%% file: texexp.tex
\pdfoutput=1
\documentclass[acmtog]{acmart}
\usepackage{booktabs} 

\setcitestyle{square}

\usepackage[ruled]{algorithm2e} 

\SetAlFnt{\small}
\SetAlCapFnt{\small}
\SetAlCapNameFnt{\small}
\SetAlCapHSkip{0pt}
\IncMargin{-\parindent}

\setcitestyle{square}

\acmJournal{TOG}
\acmVolume{37}
\acmNumber{4}
\acmArticle{49}
\acmYear{2018}
\acmMonth{8}

\setcopyright{acmcopyright}

\acmDOI{10.1145/3197517.3201286}

\input{macros}

\begin{document}
\title{Non-Stationary Texture Synthesis by Adversarial Expansion}

\author{Yang Zhou}
\authornote{Yang Zhou and Zhen Zhu are joint first authors}
\affiliation{\institution{Shenzhen University }}
\affiliation{\institution{Huazhong University of Science \& Technology}}
\author{Zhen Zhu}
\author{Xiang Bai}\affiliation{\institution{Huazhong University of Science and Technology}}
\author{Dani Lischinski}\affiliation{\institution{The Hebrew University of Jerusalem}}
\author{Daniel Cohen-Or}\affiliation{\institution{Shenzhen University}}
\affiliation{\institution{Tel Aviv University}}
\author{Hui Huang}
\authornote{Corresponding author: Hui Huang (hhzhiyan@gmail.com)}
\affiliation{\department{College of Computer Science \& Software Engineering}
	\institution{Shenzhen University}}

\renewcommand\shortauthors{Y. Zhou, Z. Zhu, X. Bai, D. Lischinski, D. Cohen-Or, and H. Huang}

\begin{abstract}
	
The real world exhibits an abundance of non-stationary textures. Examples include textures with large scale structures, as well as spatially variant and inhomogeneous textures.
While existing example-based texture synthesis methods can cope well with stationary textures, non-stationary textures still pose a considerable challenge, which remains unresolved. In this paper, we propose a new approach for example-based non-stationary texture synthesis. Our approach uses a generative adversarial network (GAN), trained to double the spatial extent of texture blocks extracted from a specific texture exemplar. Once trained, the fully convolutional generator is able to expand the size of the entire exemplar, as well as of any of its sub-blocks. We demonstrate that this conceptually simple approach is highly effective for capturing large scale structures, as well as other non-stationary attributes of the input exemplar. As a result, it can cope with challenging textures, which, to our knowledge, no other existing method can handle.

\end{abstract}

%
%


%
%


\begin{teaserfigure}
	\centering
	\includegraphics[width=0.99\textwidth]{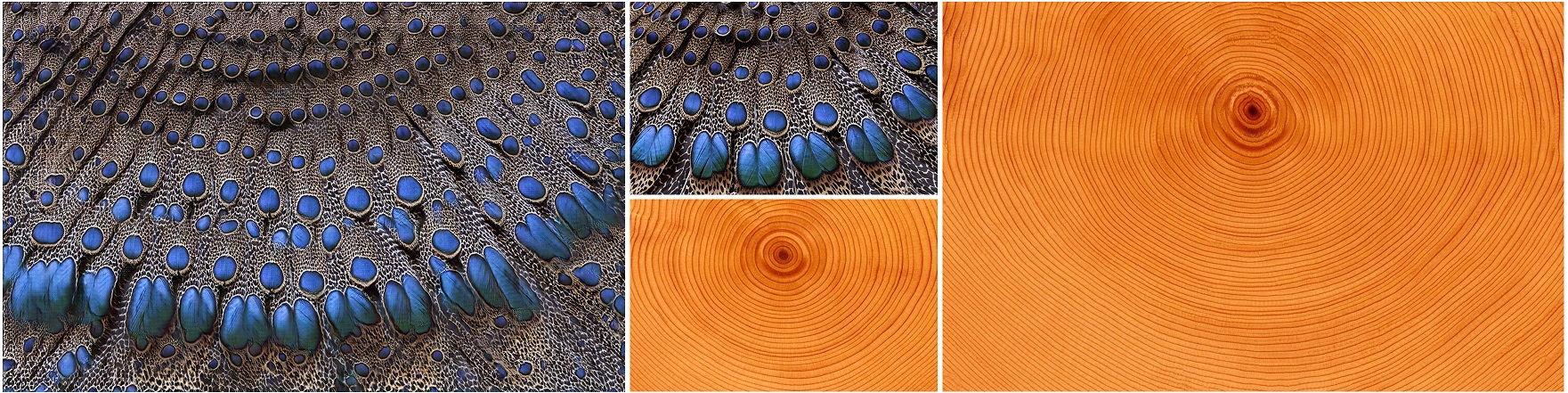}
	\caption{Examples of two extremely challenging non-stationary textures (middle column), synthesized by our method (left and right). Note that our method succeeds in reproducing and extending the global structure and trends present in the input exemplars.}
	\label{fig:teaser}
\end{teaserfigure}

\maketitle

\input{introduction}

\input{related}

\input{method}

\input{results}

\input{summary}

\section*{Acknowledgments}
We thank the anonymous reviewers for their valuable comments. This work was supported in part by NSFC (61522213, 61761146002, 61602461,  6171101466), 973 Program (2015CB352501), Guangdong Science and Technology Program (2015A030312015), the Israel Science Foundation (2366/16), the ISF-NSFC Joint Research Program (2217/15, 2472/17) and the Shenzhen Innovation Program \\
(KQJSCX20170727101233642, JCYJ20151015151249564).

\bibliographystyle{ACM-Reference-Format}
\bibliography{texexp-references}

\end{document}

%% file: macros.tex
\usepackage{color}
\definecolor{purple}{rgb}{0.65,0,0.65}
\definecolor{blue}{rgb}{0, 0.2, 0.8}
\definecolor{orange}{rgb}{0, 0.6, 0}
\definecolor{red}{rgb}{0.8, 0.2, 0.2}
\definecolor{gre}{rgb}{0.1, 0.6, 0.1}
\definecolor{magenta}{rgb}{0.5, 0.0, 1.0}
\definecolor{cyan}{rgb}{0, 0.65, 0.65}
\usepackage{xcolor}

\newcommand{\new}[1]{{\color{black}#1}}

\newcommand{\yang}[1]{{\color{black}#1}}

\newcommand{\ignorethis}[1]{}

\newcommand{\etal}{et al.}

%% file: introduction.tex
\section{Introduction}
\label{sec:intro}

\begin{figure*}[htbp]
	\includegraphics[width=0.915\textwidth]{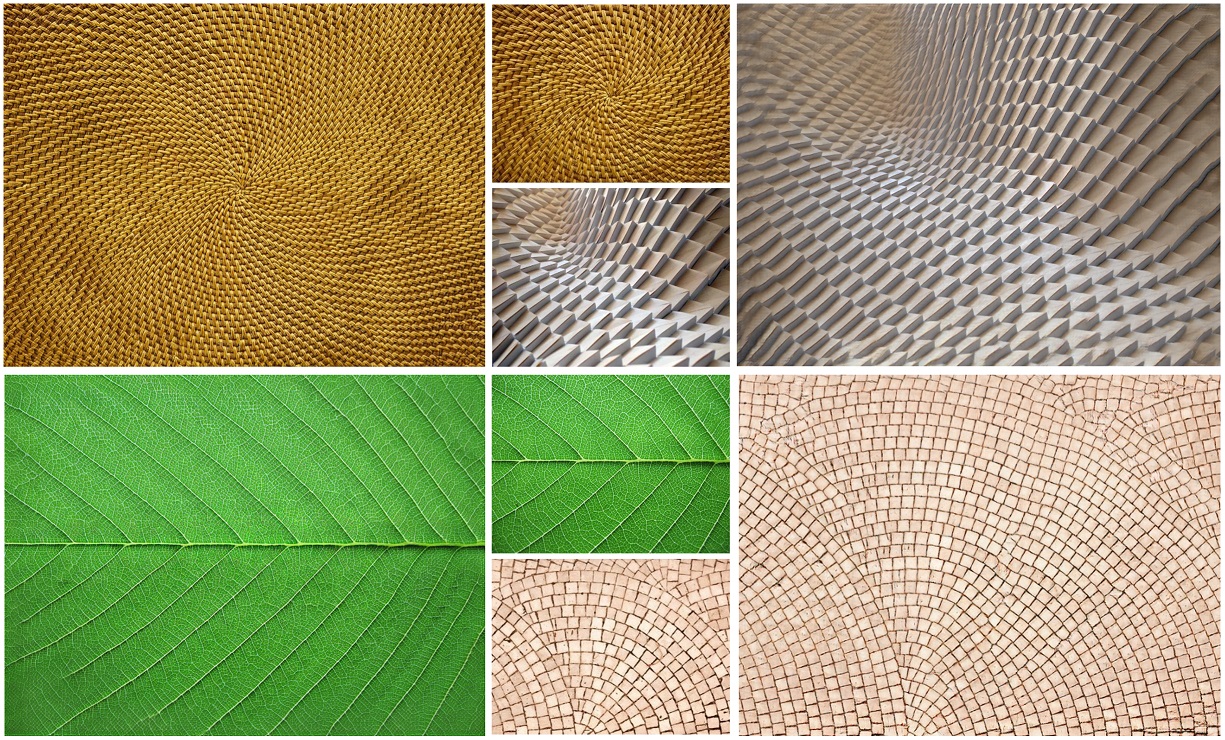}
	\caption{Four non-stationary textures (middle) and our synthesis results (left and right). Again, the global structure present in the input exemplars is preserved and extended in our results. For example, in the leaf texture, additional veins appear, whose orientation and spacing is consistent with the input.}
	\label{fig:basic_challenges}
\end{figure*}

Example-based texture synthesis has been an active area of research for over two decades.
However, despite excellent results for many classes of textures, example-based synthesis of significantly non-stationary textures remains an open problem.
Examples of non-stationary textures include textures with large-scale irregular structures, or ones that exhibit spatial variance in various attributes, such as color, local orientation, and local scale.
Inhomogeneous textures, such as weathered surfaces are another challenging example of non-stationarity. 
Several challenging non-stationary examples are shown in Figures~\ref{fig:teaser} and~\ref{fig:basic_challenges}.

During the previous decade, stitching-based \cite{Efros:2001,Kwatra2003} and optimization-based \cite{Kwatra2005,Wexler07} approaches have established themselves as highly effective for example-based texture synthesis.
More recently, deep learning based approaches for texture synthesis have begun to gain popularity. However, Figure~\ref{fig:comparisons} demonstrates that none of the existing state-of-the-art methods are able to successfully cope with significantly non-stationary input exemplars. Depending on the assumptions of each method, the results are either stationary or periodic, failing to mimic the large-scale structure and spatially variant attributes of the exemplars.


The fundamental goal of example-based texture synthesis is to generate a texture, usually larger than the input, that faithfully captures all the visual characteristics of the exemplar, yet is neither identical to it, nor exhibits obvious unnatural looking artifacts.
Given this goal, a major challenge of non-stationary texture synthesis lies in preserving the large-scale structures present in the exemplar. Consider, for example, the nearly co-centric wood rings in the right example in Figure \ref{fig:teaser}; reproducing this structure is essential for maintaining the visual similarity of the outcome to the input, and preserving the natural appearance of wood.
Additionally, it is crucial to reproduce the manner in which local patterns vary across the spatial domain, such as the changes in scale in the left example in Figure~\ref{fig:teaser}. 
These requirements are challenging for existing methods, most of which operate by enforcing similarity of local patterns and/or of certain global statistics to those of the exemplar.

In this work, we propose a new method for example-based synthesis of non-stationary textures, which uses a generative adversarial network (GAN) for this purpose.
Conceptually, our approach is, in fact, extremely simple.
The goal of the generator network is to learn how to \emph{expand} (double the spatial extent) an arbitrary texture block cropped from the exemplar, such that expanded result is visually similar to a containing exemplar block of the appropriate size. 
The visual similarity between the expanded block and an actual containing block is assessed using a discriminator network.
The discriminator is trained (in parallel to the generator) to distinguish between actual larger blocks from the exemplar and those produced by the generator.
This self-supervised adversarial training takes place for each specific texture exemplar.
Once trained, the fully convolutional generator may be used to generate extended textures up to double the original exemplar's size, that visually closely resemble the exemplar.
Even larger textures may be synthesized by feeding the generator with its own output.

Our approach also supports texture transfer: when a generator trained using a certain texture exemplar is fed with a pattern taken from another image or texture, the resulting synthesized texture follows the large scale structure from the input pattern.


At first glance, our approach may resemble deep super-resolution approaches, such as SRGAN \cite{Ledig2016}. Note, however, that super-resolution aims to enhance (sharpen) the \emph{already existing} content of an image patch or block. In contrast, our approach learns to \emph{inject} new content! This is evident in the examples of our results shown in Figures \ref{fig:teaser} and \ref{fig:basic_challenges}: all these results exhibit more elements (feathers, wood rings, leaf veins, tiles, etc.) than present in the input exemplar. Unlike in super-resolution, the size and spacing of the elements remains similar to the input, but additional elements are added without obvious repetition of the original ones.

In summary, through a variety of results and comparisons, we show that using a conceptually simple adversarial training strategy, we are able to cope with an unprecedented array of highly non-stationary textures, which to our knowledge none of the currently existing methods are able to handle.

%% file: related.tex
\section{Related work}
\label{sec:related}

We begin with a brief review of classical example-based texture synthesis methods, followed by a more detailed discussion of recent deep learning based approaches.
In either category, the existing methods are unable to cope with highly inhomogeneous textures, or textures that exhibit large scale or global structures.

\subsection{Classical approaches}

Example-based texture synthesis has been extensively researched for over twenty years, and we refer the reader to Wei \etal~\shortcite{Wei2009} for a comprehensive survey.
The most effective approaches have been non-parametric methods, which include pixel-based methods \cite{Efros99,Wei2000}, stitching-based methods \cite{Efros:2001,Kwatra2003}, optimization-based methods \cite{Kwatra2005,Wexler07}, and appearance-space texture synthesis \cite{Lefebvre2006}.

Image melding \cite{Darabi12} unifies and generalizes patch-based synthesis and texture optimization, while Kaspar \etal~\shortcite{Kaspar:2015} describe a self-tuning texture optimization approach, which uses image melding with automatically generated and weighted guidance channels.
These guidance channels are designed to help reproduce the middle-scale structures present in the texture exemplar.
However, as demonstrated in Figure \ref{fig:comparisons}, this state-of-the-art classical approach is unable to capture and reproduce the large-scale or global structure that may be present in the exemplar.

In general, while classical non-parametric methods are typically able to reproduce small scale structure, they assume a stationary Markov Random Field (MRF) model, making it difficult for them to cope with highly inhomogeneous textures, which violate this assumption.
Thus, control of large scale structure and inhomogeneity has typically required user-provided or automatically generated guidance maps (e.g., \cite{Hertzmann01,Zhang2003,Rosenberger:2009,Zhou2017}).
We are not aware of \emph{any} classical example-based texture synthesis method capable of automatically coping with challenging non-stationary exemplars, such as the ones shown in Figures~\ref{fig:teaser} and \ref{fig:basic_challenges}.

\begin{figure}[tb]
	\includegraphics[width=0.99\linewidth]{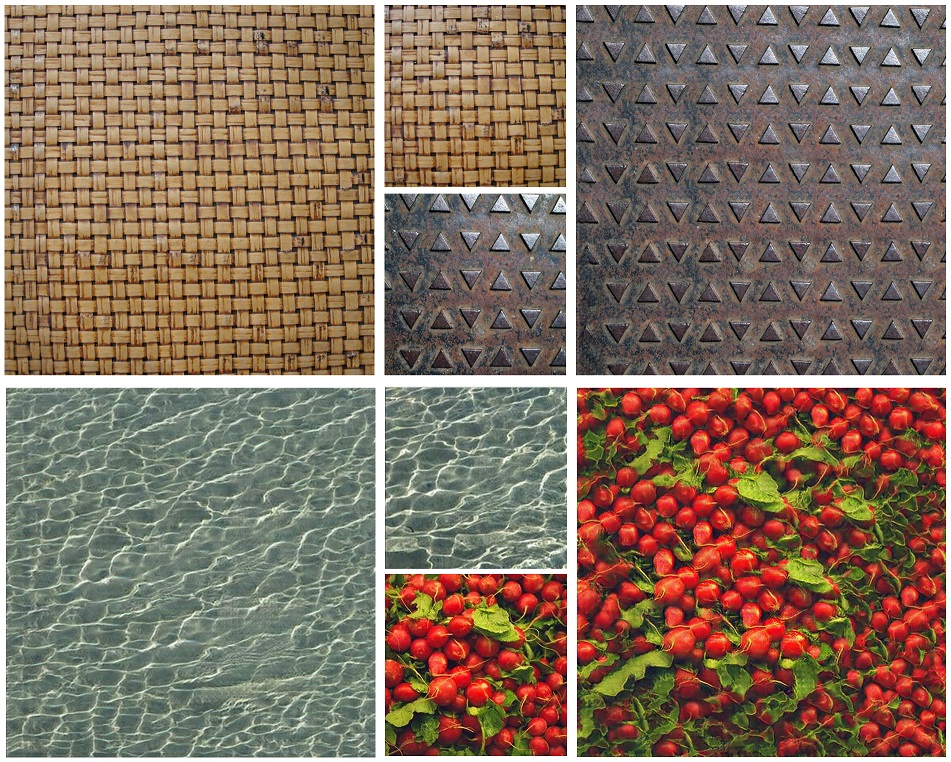}
	\caption{Our method can also be successfully applied to stationary, regular, near-regular, or stochastic textures.}
	\label{fig:basic_classic}
\end{figure}

Certain classes of global structures can be handled by classical texture synthesis approaches.
For example, Liu et al.~\shortcite{Liu04} analyze near-regular textures and explicitly model their geometric and photometric deviations from a regular tiling. In contrast, our approach does not make any assumptions regarding the structure, nor does it attempt to analyze it. Yet, with the same deep architecture and training strategy, we are also able to synthesize regular and near-regular textures, as demonstrated in Figure \ref{fig:basic_classic}.

\subsection{Deep Learning based approaches}

Gatys et al.~\shortcite{Gatys-texture-2015} were, to our knowledge, the first to use a deep neural network for example-based texture synthesis.
They characterize an input texture by a collection of Gram matrices, each defined by inner products between feature channels at a certain convolution layer of a pre-trained image classification network (in their case VGG-19 \cite{Simonyan2014}).
An image input to the network is then iteratively optimized (using back-propagation) so as to minimize a loss function defined as a sum of weighted differences between its Gram matrices and those of the original exemplar.

The loss function of Gatys et al.~\shortcite{Gatys-texture-2015}, often referred to as \emph{Gram loss} or \emph{style loss} (in the context of neural style transfer \cite{Gatys2015c}), is unable to capture well regularities and larger structures in the texture.
Sendik and Cohen-Or \shortcite{Sendik2017} address this deficiency by introducing \emph{structural energy}, based on deep inter-feature correlations.
This approach considerably improves synthesis quality for textures with regular structure, but still can not deal with non-stationary structures. 

Gatys et al.~\shortcite{Gatys2015c} extend their Gram-based texture synthesis approach to perform artistic style transfer.
To achieve this, a \emph{content loss} term is added to the Gram-based style loss.
This approach still requires iterative optimization to obtain each result.
Ulyanov et al.~\shortcite{Ulyanov2016} and Johnson et al.~\shortcite{Johnson2016} both propose a fast implementation of Gatys et al.'s texture synthesis and style transfer using a single feed-forward pass through a network trained for a specific texture (style).
The idea is to move the computational burden to the learning stage: a generator network is trained by using a pre-trained \emph{descriptor network} (also referred to as \emph{loss network}) based on VGG-19 in place of a loss function.
The quality of the results is comparable to Gatys et al., while the synthesis itself is extremely fast, once the network has been trained. In Figure \ref{fig:comparisons} we compare our results to those of Ulyanov et al.~(TextureNets), which can also be viewed as a comparison with Gatys et al.~\shortcite{Gatys2015c} and Johnson et al.~\shortcite{Johnson2016}.



Several works have utilized Generative Adversarial Networks (GANs) to perform texture synthesis and style transfer. 

Li and Wand \shortcite{Li2016} introduce Markovian Generative Adversarial Networks (MGANs).
Rather than capturing style with global statistics, their generator network is trained using a discriminator which examines statistics of Markovian neural patches, i.e., local patches on feature maps extracted by a pre-trained VGG network, thereby imposing a Markov Random Field prior.
As in other style transfer approaches, explicit layout constraints may also be imposed via a ``content'' image provided as additional input.

Jetchev et al.~\shortcite{Jetchev2016} also utilize GANs for texture synthesis, where texture patches are generated from random noise, and patches of the same size as the generated output are randomly selected from the exemplar as the ground truth for adversarial training. However, their method failed to produce high quality textures consistently.
Bergmann et al.~\shortcite{Bergmann2017} extend this approach by introducing a periodic function into the input noise, which enables synthesizing periodic textures with high quality. However, the approach, referred to as PSGAN, is limited to periodic textures and thus is not applicable to most real-world textures, as demonstrated in Figure~\ref{fig:comparisons}.


Isola et al.~\shortcite{pix2pix2016-arxiv} demonstrate the effectiveness of GANs for a variety of image-to-image translation tasks.
Zhu et al.~\shortcite{CycleGAN2017} introduce CycleGANs, where the translation network can be trained with unpaired training data.
In these tasks, the input and output differ in appearance, but correspond to different renderings of the same underlying structure.
This is not the case in our approach, where the goal is to extend the global structure of the exemplar. We do so by introducing new instances of local patterns, which are similar, but not identical, to those present in the exemplar.




%% file: method.tex
\section{Our Approach}
\label{sec:method}

We begin this section with an overview of our approach, followed by a more detailed explanation of the network architectures used and the training procedure.

\begin{figure}[t]
	\includegraphics[width=\columnwidth]{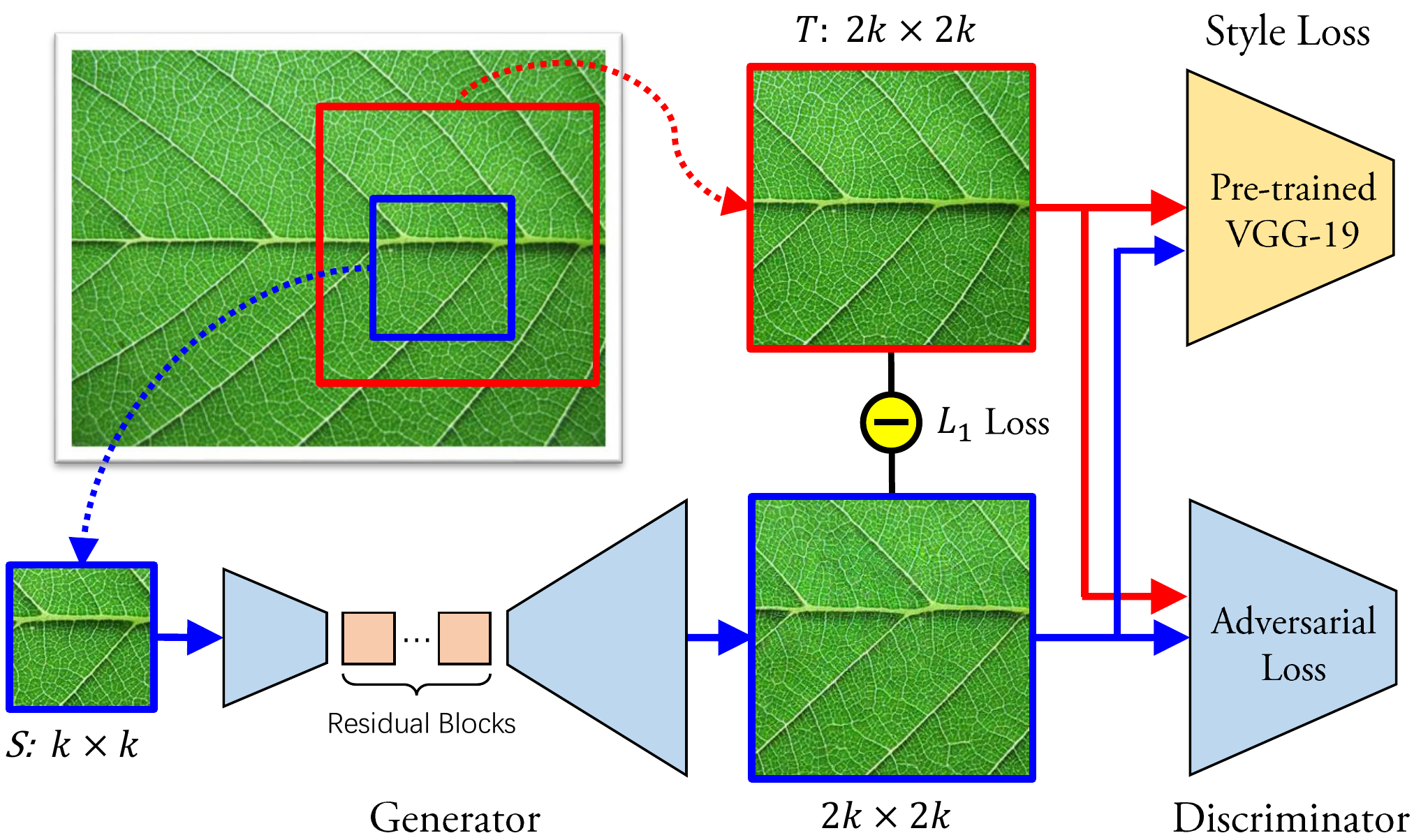}
	\caption{Method overview. The generator learns to expand $k \times k$ texture blocks into $2k \times 2k$ ones using a combination of adversarial loss, $L_1$ loss and style loss.}
	\label{fig:overview}
\end{figure}

Our approach is very simple conceptually: given that our ultimate goal is to generate larger instances that perceptually resemble a smaller input texture exemplar, the main idea is to teach a fully convolutional generator network how to do just that.
The approach is depicted by the diagram in Figure~\ref{fig:overview}.
More specifically, given a $k \times k$ \emph{source block} $S$ cropped from the input exemplar, the generator must learn to produce a $2k \times 2k$ output, which is perceptually similar to an enclosing \emph{target block} $T$ of the latter size.
Note that this training procedure is self-supervised: the ground truth extended texture blocks are taken directly from the input texture.
Since the generator is a fully-convolutional network, once it has been trained, we can apply it onto the entire input exemplar, or a sufficiently large portion thereof, to generate a texture that is larger than the input (up to double its size).

It is well known that pixel-based metrics, such as $L_1$ or $L_2$ are not well suited for assessing the perceptual differences between images.
This is even more true when the goal is to compare different instances of the same texture, which are the output of texture synthesis algorithms.
On the other hand, recent work has shown the effectiveness of adversarial training and GANs for a variety of image synthesis tasks \cite{pix2pix2017-cvpr,Ledig2016,CycleGAN2017}, including texture synthesis \cite{Li2016,Bergmann2017}.
Thus, we also adopt an adversarial training approach to train our generator. 
In other words, our generator $G$ is trained alongside with a discriminator $D$ \cite{Goodfellow2014}.
The discriminator $D$ is trained to classify whether a $2k \times 2k$ texture block is real (a crop from the input exemplar) or fake (synthesized by $G$).

In our approach, a dedicated GAN must be trained for each input exemplar, which takes considerable computational resources.
But once the fully-convolutional generator has been trained, large texture blocks may be synthesized from smaller ones in a single forward pass through the network, which is extremely fast when the network runs on the GPU.
The size of the $k \times k$ source blocks that we use during the training stage should be chosen large enough to capture the non-stationary behavior across the input exemplar.
On the other hand, it should be small enough relative to the size of the exemplar, so that we can extract a sufficient number of different $2k \times 2k$ target blocks to train the network.
In our current implementation we set $k = 128$, and our exemplars are typically of size $600\times 400$. 

\begin{figure}[t]
	\includegraphics[width=\columnwidth]{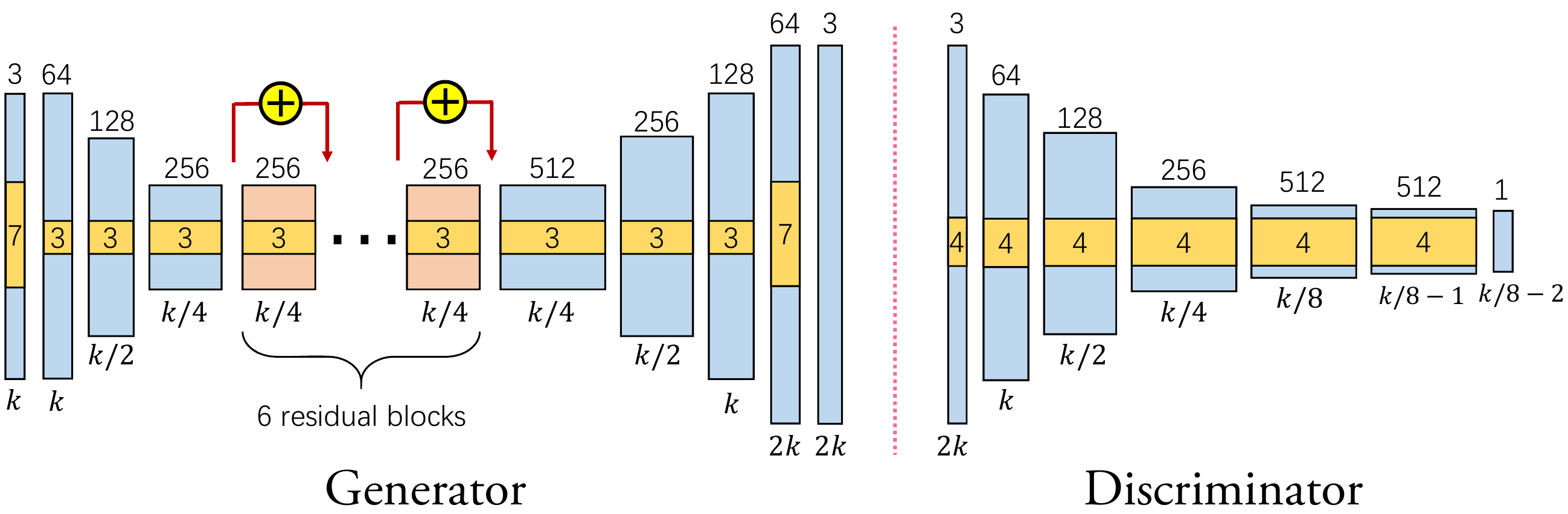}
	\caption{Architecture of our generator and discriminator. The number of feature channels and the spatial resolution of feature maps are respectively specified on the top of and under each block, while the kernel sizes are specified in the central yellow regions.
	}
	\label{fig:architecture}
\end{figure}

\begin{figure*}[t]
	\includegraphics[width=\textwidth]{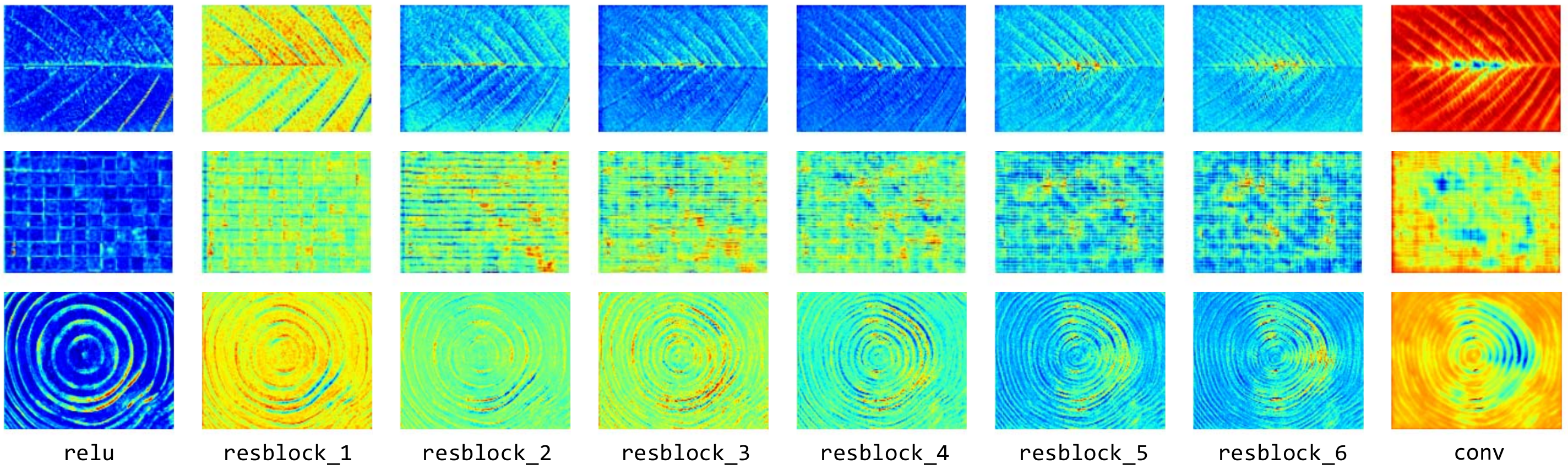}
	\caption{Visualization of feature maps output by the middle part of our generator. Besides the intermediate results of residual blocks (from \texttt{resblock\_1} to \texttt{resblock\_6}) we also visualize the final output of encoding stage (\texttt{relu}), and the feature map output by the first convolution layer of the decoder (\texttt{conv}). While the leaf (top), bricks (middle) and wood ring (bottom) textures have very different large-scale structures, it may be observed that all of the new structures emerge in the course of the residual block chain. The creation of new structures is typically complete before the end of the chain, as evidenced by the similarity between the \texttt{resblock\_5} and \texttt{resblock\_6} feature maps.
	}
	\label{fig:visualization}
\end{figure*}

\subsection*{Network architecture}

As explained earlier, we would like to model the generator as a fully-convolutional deep neural network. Using a fully-convolutional network allows us to apply the generator to arbitrary-sized inputs at test time, and reduces the number of parameters, compared to networks with fully connected layers.
Network depth is important both for the expressive power of the generator, and for increasing the receptive field of the network's neurons.
Since our goal is to capture large-scale non-stationary behavior across the source texture block, at some level of the network the receptive field should approach the size of the source block. This may be effectively achieved by introducing a chain of residual blocks \cite{He2016}.

A generator architecture that satisfies our requirements was, in fact, already proposed by Johnson et al.~\shortcite{Johnson2016}, who demonstrated its effectiveness for neural style transfer and for super-resolution. The same generator was later successfully used by Zhu et al.~\shortcite{CycleGAN2017} for a variety of unpaired image-to-image translation tasks.
Neural style transfer is closely related to texture synthesis; thus, we adopt a similar architecture for our generator.

The architecture of the generator is shown in the diagram in Figure \ref{fig:architecture}.
The network consists of three convolution layers, two of which use stride-2 convolutions that reduce the spatial dimensions of the input.
These three layers are followed by a sequence of six residual blocks \cite{He2016}.
The receptive field of the neurons at the end of the residual chain is $109 \times 109$, i.e., close to the size of our training source blocks.
From this point onward, we first double the number of channels, after which the spatial dimensions are doubled three times via 
strided deconvolution layers, yielding twice the original spatial resolution. Finally, the multi-channel result of this process is combined back into three image channels.
Similarly to previous work we use batch normalization after each convolution, except the last one.

Figure~\ref{fig:visualization} visualizes the feature maps output by the residual blocks of our generator. Through this visualization, we can gain a better understanding of how the generator works. The different activation maps after the downsampling stages (\texttt{relu}) reveal that they encode details at various scales and orientations. No new large scale structures appear to be present yet. The situation is different by the end of the residual chain (\texttt{resblock\_6}), where we can see that the number of the large scale structures (leaf veins, bricks or wood rings) has roughly doubled.
Thus, the residual blocks appear to be responsible for introducing new large scale structures. This makes a lot of sense, since each residual block is capable of spatially transforming its input (via its two convolution layers), and adding the transformed result to its input. It appears that a chain of such blocks is capable of learning which structures, among those present in the chain's input, should be replicated, and how the resulting replicas should be spatially transformed before they are recombined with the original pattern. For example, for the leaf texture, it is capable of learning that the leaf vein structures should be shifted horizontally after replication, while for the wood rings texture it learns to shift the replicated rings radially. In either case, the amount of large scale structure is roughly doubled. However, when a generator trained on a certain texture is applied to an input consisting of completely different structures, these structures are not replicated, as demonstrated by the results in Figure \ref{fig:style}.


While Johnson et al.~\shortcite{Johnson2016} employ a loss network, which is used to compute the style loss and content loss functions of Gatys et al.~\shortcite{Gatys2015c}, we require a loss function with more sensitivity to spatial arrangement of texture elements and their spatially variant appearance.
Thus, we adopt the PatchGAN discriminator \cite{Ledig2016,Li2016,pix2pix2017-cvpr,CycleGAN2017} instead. The discriminator architecture is shown in Figure \ref{fig:architecture} (bottom right).
This fully-convolutional network halves the spatial resolution of the input four times, while doubling the number of channels.
The neurons at the sixth layer may be viewed as texture descriptors of length 512, representing overlapping texture patches of size $142 \times 142$ in the input.
Each of these 512-dimensional descriptors is then projected into a scalar (using a $1 \times 1$ convolution, followed by a sigmoid), and the resulting 2D pattern is classified as real or fake using binary cross-entropy. 


\subsection*{Training procedure}

Our training process follows the one outlined in the pioneering work of Goodfellow et al.~\shortcite{Goodfellow2014}: we repeatedly alternate between performing a single training iteration on the discriminator $D$, and a single training iteration on the generator $G$. In each training iteration, we randomly select one $256 \times 256$ target block $T$ from the exemplar to serve as the ground truth, as well as a random $128 \times 128$ source block $S$, contained in $T$, which is fed as input to the generator.
\new{
For maximum utilization of the available data we choose to not set aside a validation or a test set. Nevertheless, our results show that the network is able to plausibly expand unseen input texture blocks that are different in both size and content from those encountered during training, and it is easy to see that it does not merely memorize patterns. It is also capable of texture transfer, as demonstrated in Figure \ref{fig:style}.
}

In addition to the standard adversarial loss function \cite{Goodfellow2014} $\mathcal{L}_{\mathit{adv}}$, we use two additional loss terms: $L_1$ loss $\mathcal{L}_{L_1}$ and style loss $\mathcal{L}_{\mathit{style}}$ \cite{Gatys-texture-2015}:
\begin{equation}
\label{eq:loss}
\mathcal{L}_{\mathit{total}} = \mathcal{L}_{\mathit{adv}} + \lambda_1 \mathcal{L}_{L_1} + \lambda_2 \mathcal{L}_{\mathit{style}}, 
\end{equation}
where $\lambda_1 = 100$ and $\lambda_2 = 1$. As we shall
demonstrate in our ablation study in Section \ref{sec:selfeval}, the adversarial loss appears to be the main workhorse, while the other two terms help stabilize the training and slightly reduce artifacts. 

Following Gatys et al. \shortcite{Gatys-texture-2015}, we compute the style loss using a pre-trained (on ImageNet) VGG-19 model, and compute Gram matrices for the ReLU activated feature maps output by the \texttt{relu1\_1}, \texttt{relu2\_1}, \texttt{relu3\_1}, \texttt{relu4\_1}, and \texttt{relu5\_1} layers.
The weights used to sum up the corresponding Gram losses are set to 0.244, 0.061, 0.015, 0.004, and 0.004, respectively. More specifically, they are given by 1000/(64 x 64), 1000/(128 x 128), 1000/(256 x 256), 1000/(512 x 512), and 1000/(512 x 512).

We choose Adam \cite{Kingma2014-Adam} as our optimization method with momentum set to 0.5, and train our models for up to 100,000 iterations. Learning rate is set to 0.0002 initially and kept unchanged for the first 50,000 iterations. Then, the learning rate linearly decays to zero over the remaining 50,000 iterations. Weights of convolutional layers are initialized from a Gaussian distribution with mean 0 and standard deviation 0.02. We train and test all our models on an NVIDIA Titan Xp GPU with 12GB of GPU memory.

%% file: results.tex
\section{Results}
\label{sec:results}

Our approach was implemented using PyTorch, building on publicly available existing implementations of its various components.
Generators were trained for a variety of input exemplars of sizes around 600$\times$400 pixels.
Training our GAN on an exemplar of this size takes about 5 hours for 100,000 iterations on a PC equipped with a NVIDIA Titan Xp GPU with 12GB memory.
In many cases the results no longer improve after around 36,000 iterations (under 2 hours).  
\yang{Our implementation, as well as our trained models and other supplementary materials, are all available on the project page\footnote{http://vcc.szu.edu.cn/research/2018/TexSyn}.}

Once the generator has been trained it takes only 4--5 milliseconds to double the size of a 600$\times$400 texture, since this requires only a single feed-forward pass through the generator.

A few of our synthesis results from challenging non-stationary texture exemplars exhibiting irregular large-scale structures and inhomogeneities are shown in Figures~\ref{fig:teaser} and \ref{fig:basic_challenges}. 
In all of these examples, the global structure present in the input exemplars is successfully captured and extended by our method.
Of course, our method is also applicable to more stationary textures as well, including textures with regular, near-regular, or stochastic structures.
Four examples of our results on such textures are shown in Figure \ref{fig:basic_classic}.
Results for additional textures are included in the supplementary material.


\input{figures/comparisons/comparisons}

\subsection{Comparison}

Figure \ref{fig:comparisons} compares our results with those produced by a number of state-of-the-art methods. The first column shows the input exemplars, which include both non-stationary and stationary textures. Our results are shown in the second column. The third column shows results produced by self-tuning texture optimization \cite{Kaspar:2015}, which is a representative of classical optimization-based texture synthesis methods. 
The next four columns show results produced by several recent deep learning based approaches: TextureNets by Ulyanov et al.~\shortcite{Ulyanov2016}, a feed-forward version of the method proposed by Gatys et al.~\shortcite{Gatys-texture-2015}; DeepCor by Sendik and Cohen-Or~\shortcite{Sendik2017} improves upon Gatys et al.'s approach by introducing a deep correlations loss that enables better handling of large scale regular structures; MGANs of Li and Wand \shortcite{Li2016}, the first texture synthesis method to use adversarial training, employing a discriminator that examines statistics of local patches; and PSGAN of Bergmann et al. \shortcite{Bergmann2017}, which learns to convert periodic noise into texture patches sampled from the exemplar.
 
These comparisons demonstrate that our approach is able to handle large-scale non-stationarity much better than existing methods, while for stationary or homogeneous textures, we produce comparable results to the state-of-the-art approaches. Additional comparison results are contained in our supplementary materials.

In terms of computation times, the self-tuning method \cite{Kaspar:2015} takes about 20 minutes per result; the deep learning based methods take between 1 hour of training per exemplar with TextureNets \cite{Ulyanov2016}, to 
12 hours of training an PSGAN \cite{Bergmann2017}, and up to 8 hours for each result using Deep Correlations \cite{Sendik2017}. Thus, while the training time of our method is much slower than the time it takes to synthesize a single texture with a classical method, it is far from being the slowest among the deep-learning based methods.  

\subsection{Diversification}
\label{subsec:variability}

It is important for a texture synthesis algorithm to be able to produce a diverse collection of results from a single input exemplar.
Since our method does not generate textures from a random seed or noise, we have explored a number of alternatives for diversifying the output.
The simplest approach is to simply feed different subwindows of the exemplar as input to be expanded by our generator.
Since the appearance across non-stationary exemplars varies greatly, cropping and expanding different windows may result in quite different results.
This is demonstrated in Figure \ref{fig:basic_variation}, which shows two different 512$\times$512 synthesis results for each exemplar, obtained by taking two random 256$\times$256 crops as input. 

	

\input{figures/variation_crops/different_crops}

\input{figures/variation_tile/tile_shuffle}

For exemplars with a more stochastic and stationary nature, without a clear global structure, it is also possible to diversify the results by reshuffling or perturbing the source texture.
Specifically, for sufficiently stationary textures, we have been able to produce a wide variety of synthesis results by reshuffling the exemplar's content. Figure~\ref{fig:variation_tile} shows three exemplars, each of which was split into 4$\times$4 tiles, which were randomly reshuffled each time before feeding into the generator to yield different results.
We have also experimented with adding Perlin noise to both stationary and non-stationary exemplars.
We found that the changes among different results generated in this manner are more moderate, and are best presented using animated sequences; we include a number of such animations in our supplementary materials.


\input{figures/ablation/ablation}
\input{figures/patchGAN/patchGAN}

\subsection{Self evaluation}
\label{sec:selfeval}

\paragraph{Ablation study.}
Figure \ref{fig:ablation} shows the results of an ablation study that we carried out in order to verify the importance of the various loss terms in Equation \ref{eq:loss}. \yang{We first train the generator with the adversarial loss switched off, i.e., without adversarial training. In this case, the generator fails to properly expand the input texture: no new large scale structures are introduced in the leaf example, and the smaller scale structures are not reproduced faithfully. Next, we turn on adversarial training and experiment with different combinations of the other two loss terms, including: adversarial loss only, adversarial and $L_1$ loss, adversarial and style loss, and the combination of all three terms.}
The visual differences between results achieved using these different combinations are quite subtle.
Clearly, the adversarial loss plays a \emph{key role} in our approach, as it alone already produces good results. 
Nevertheless, some noise and artifacts are present, which are reduced by adding the $L_1$ loss.
However, this also causes oversmoothing of local details in some areas.
In contrast, style loss enhances details, but at the same time introduces artifacts into the structures and causes some color distortions. The combination of all three terms, yields the best results, in our experience.


\paragraph{Discriminator patch size.}
The PatchGAN discriminator used in our approach is fully convolutional. Thus, it can be adjusted to examine texture patches of different sizes by changing the number of its resolution-reducing convolutional levels. We experimented with PatchGANs of six different sizes (ranging from 16 to 574). Results for two textures are shown in Figure \ref{fig:patchGAN}.
Our results on these and other textures consistently indicate that the best texture expansions are obtained using a 142$\times$142 PatchGAN.


\input{figures/stress1_crop/crop_expansion}

\input{figures/stress2_extreme/extreme}

\paragraph{Synthesis stability.}
Kaspar et al.~\shortcite{Kaspar:2015} proposed an interesting \emph{stress test} to evaluate the stability of a synthesis algorithm, which consists of feeding an algorithm with its own output as the input exemplar.
Since our approach doubles the size of its input at every stage, we conducted a modified version of this test, where after each synthesis result is obtained, we randomly crop from the result a block of the same size as the original input and feed it back to our method.
Note that we keep applying the same generator, without any re-training or fine-tuning.
Figure \ref{fig:stress1_crop} shows the results of five synthesis generations on two textures.
Obviously, since in this process we essentially zoom-in on a portion of the original texture, the global structure changes accordingly.
However, it may be seen that the smaller scale texture elements remain sharp and faithful to their shapes in the original exemplar.


\begin{figure*}[htbp]
	\includegraphics[width=1.0\textwidth]{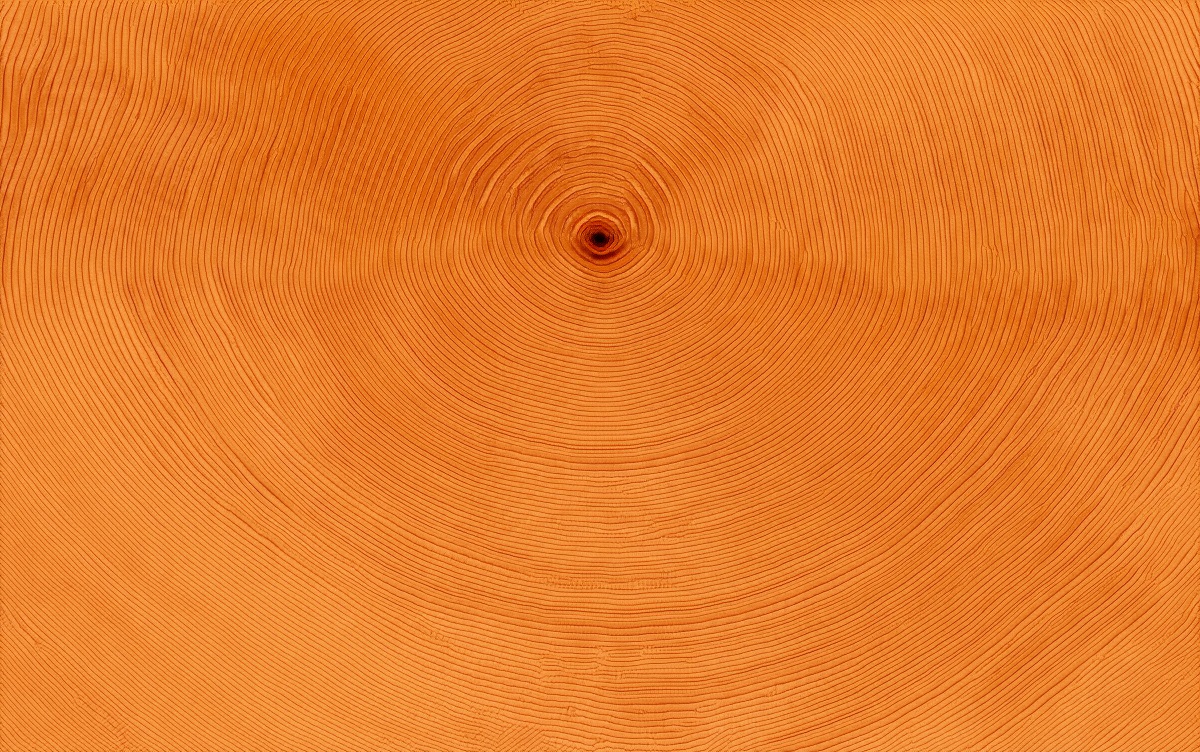}
	\caption{Expansion of the wood ring texture to a size four times larger than the exemplar by repeating the expansion twice, resulting in a 2400$\times$1504 texture. The synthesis adds additional wood rings while preserving their local appearance, as well as their global radial structure.}
	\label{fig:wood_ring_huge}
\end{figure*}

\paragraph{Extreme expansion.} 
\vspace{-1em}
Given that our method can expand the source texture up to twice its size, by repeating the expansion one can synthesize very large results.
Figure~\ref{fig:wood_ring_huge} shows the result of expanding the wood rings exemplar by a factor of four (by expanding once more the result shown in Figure \ref{fig:teaser} using the same trained model).
The result successfully maintains the radial structure of the wood rings.
Figure~\ref{fig:stress2_extreme} shows a more extreme expansion result, where starting from a 64$\times$64 patch, it is expanded to x32 of its original size via five expansion cycles. All of the cycles use the same model trained on the original exemplar.
Two additional multi-cycle expansion examples can be seen in our supplementary materials.

\begin{figure*}[htbp]
	\includegraphics[width=1.0\textwidth]{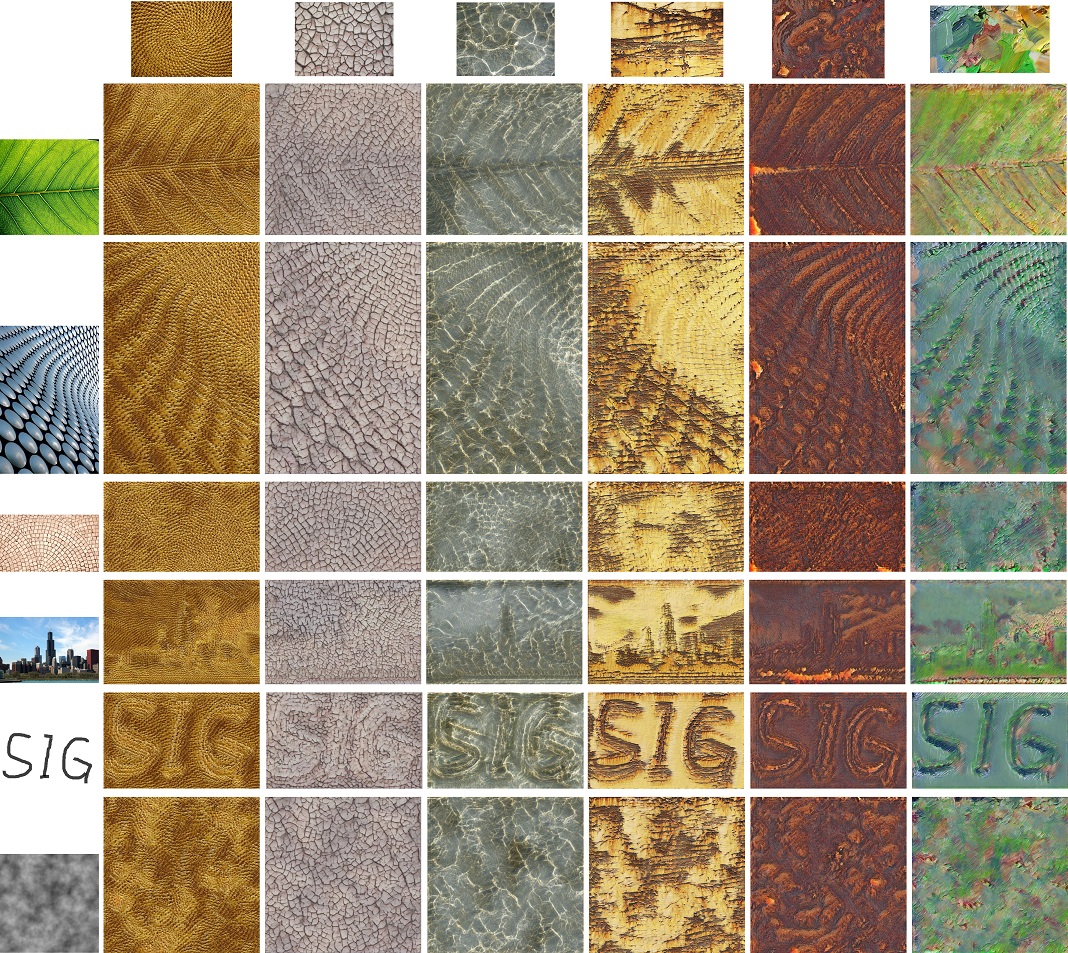}
	\caption{Texture transfer. By feeding generators trained using the texture exemplars in the top row with guiding images in the leftmost column we synthesize textures that adapt to the large scale structures present in the guiding images. \yang{Note that we can even input a simple user sketch or pure random noise (Perlin noise) and generate satisfactory results as shown in the last two rows.} }
	\label{fig:style}
\end{figure*}

\subsection{Texture Transfer}

Texture transfer is a process where a given example texture is applied to another image, guided by various properties of the latter.
Early work \cite{Efros:2001,Hertzmann01} performed texture transfer based on the brightness of the target image.
Artistic style transfer \cite{Gatys2015c} may be viewed as texture transfer guided by a more sophisticated analysis of the target's content.
Our approach, may be used without any modification to produce synthesized textures that follow the large scale structure of a \emph{guiding image}. This is achieved simply by feeding the guiding image as input to a trained generator. A collection of texture transfer results is shown in Figure~\ref{fig:style}.
The resolution of these results is twice that of the guiding images. In this case, however, no new large scale structures are produced, since the generator was not trained to extend the structures present in the guidance image.
Since our generator is not trained to extract high-level semantic information from the input image, we find that this approach is not well suited for artistic style transfer. However, Figure~\ref{fig:style} demonstrates its usefulness for synthesis of textures that follow a certain large-scale pattern.

%% file: figures/comparisons/comparisons.tex
\begin{figure*}[htbp]
	\includegraphics[width=\textwidth]{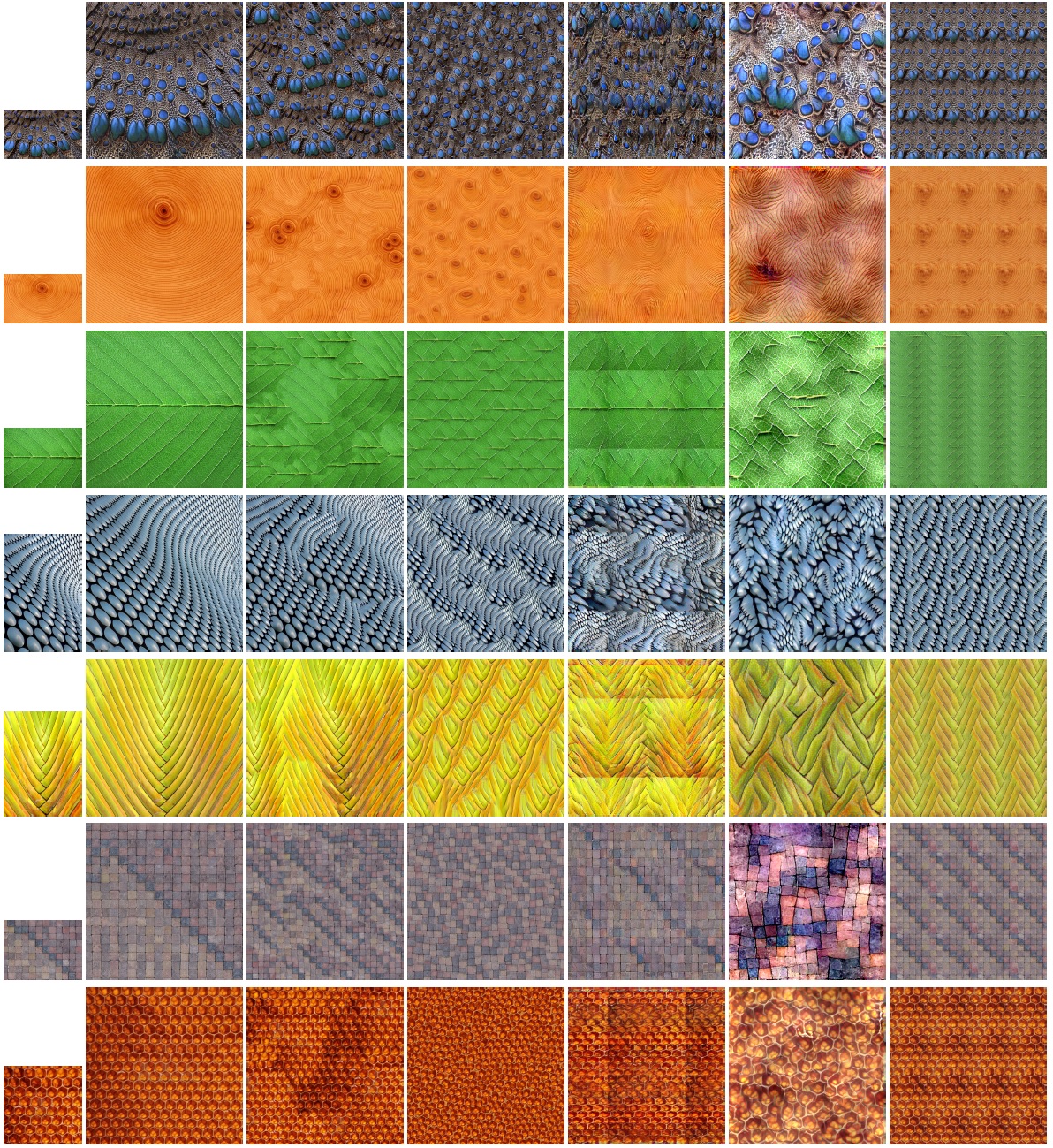}\\
	%
	\parbox{0.075\textwidth}{\centering Input}~%
	\parbox{0.15\textwidth}{\centering Our Result}~%
	\parbox{0.15\textwidth}{\centering Self-tuning}~%
	\parbox{0.15\textwidth}{\centering TextureNets}~%
	\parbox{0.15\textwidth}{\centering DeepCor}~%
	\parbox{0.15\textwidth}{\centering MGANs}~%
	\parbox{0.15\textwidth}{\centering PSGAN}%
	\caption{Comparisons to several state-of-the-art texture synthesis methods. For each texture, the results from left to right are respectively produced by our method, Self-tuning of Kaspar et al.\protect{\shortcite{Kaspar:2015}}, TextureNets of Ulyanov et al.\protect{\shortcite{Ulyanov2016}}, DeepCor of Sendik and Cohen-Or~\protect{\shortcite{Sendik2017}}, MGANs of Li and Wand\protect{\shortcite{Li2016}}, and PSGAN of Bergmann et al.~\protect{\shortcite{Bergmann2017}}.}
	\label{fig:comparisons}
\end{figure*}

%% file: figures/variation_crops/different_crops.tex
\begin{figure*}[htbp]	
	\includegraphics[width=\textwidth]{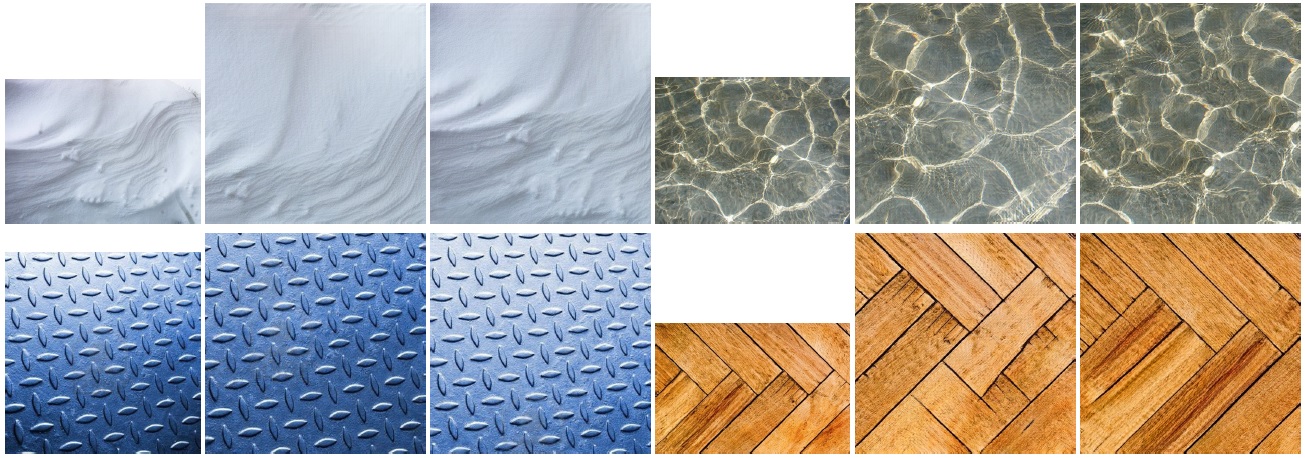}
	\caption{Diversification by cropping. For each source texture (left in each triplet), we randomly crop two 256$\times$256 sub-regions from the source texture after training, to generate different expansion results on size 512$\times$512.}
	\label{fig:basic_variation}
\end{figure*}

%% file: figures/variation_tile/tile_shuffle.tex
\begin{figure*}[htbp]	
	\includegraphics[width=\textwidth]{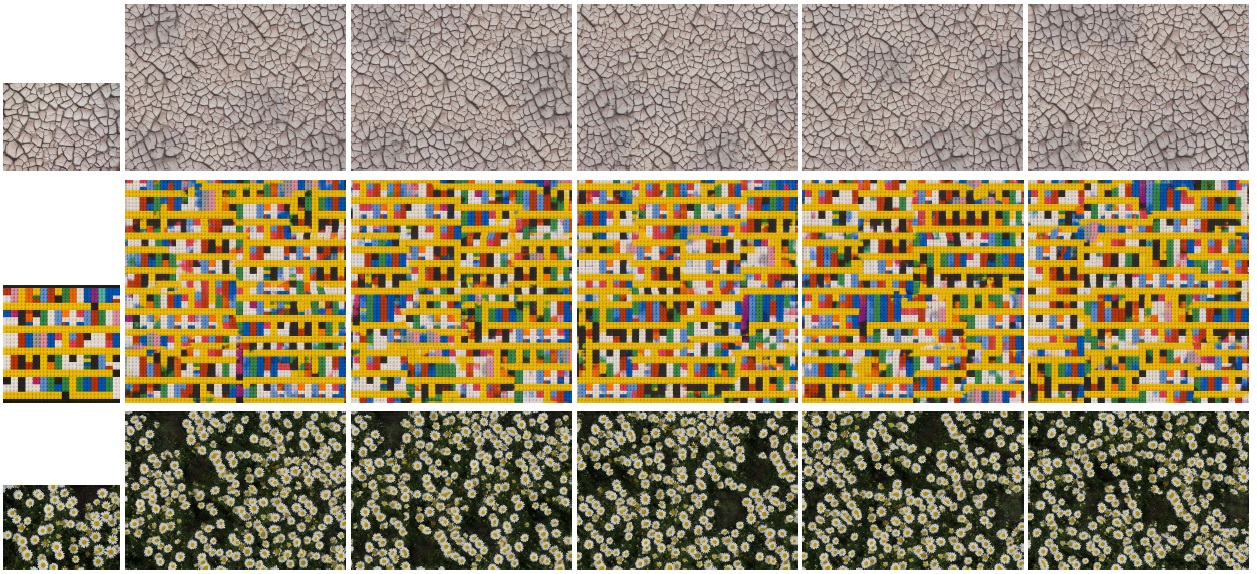}
	\caption{Diversification by tile shuffling. The exemplar used to train the generator (leftmost column) is divided into tiles, which are randomly reshuffled before feeding into the generator, yielding different results.}
	\label{fig:variation_tile}
\end{figure*}

%% file: figures/ablation/ablation.tex
\begin{figure*}[htbp]
	\centering
	\includegraphics[width=0.91\textwidth]{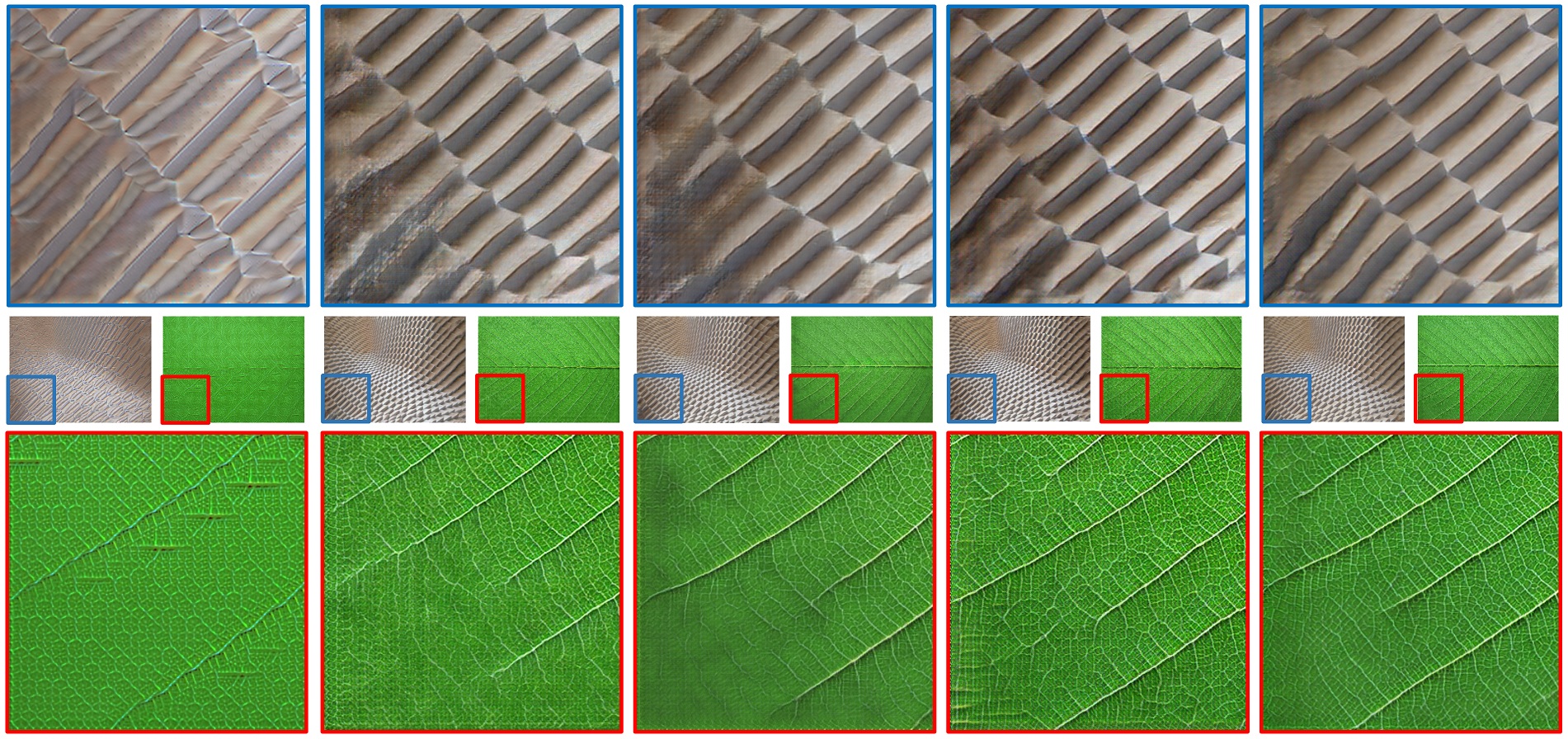} 
	\parbox{0.2\textwidth}{\rightline{$\mathcal{L}_{L_1} + \lambda_2 / \lambda_1 \mathcal{L}_\mathit{style}$}}~%
	\parbox{0.2\textwidth}{\centering{$\mathcal{L}_\mathit{adv}$} only}~%
	\parbox{0.2\textwidth}{\centering{$\mathcal{L}_\mathit{adv} + \lambda_1 \mathcal{L}_{L_1}$}}~%
	\parbox{0.16\textwidth}{\leftline{$\mathcal{L}_\mathit{adv} + \lambda_2 \mathcal{L}_\mathit{style}$}}~%
	\parbox{0.24\textwidth}{\leftline{$\mathcal{L}_\mathit{adv} + \lambda_1 \mathcal{L}_{L_1} + \lambda_2 \mathcal{L}_\mathit{style}$}}%
	\caption{Ablation study \yang{on two textures shown in Figure~\ref{fig:basic_challenges}. The leftmost column shows the expansion results without adversarial training, and the remaining columns show results of using different combinations of loss terms with adversarial loss switched on. The full results of adversarial expansion are shown in the middle row, while the top \& bottom rows zoom into the blue and red framed windows indicated in the middle row. For high-resolution full image results, please refer to our supplementary materials.} }
	\label{fig:ablation}
\end{figure*}

%% file: figures/patchGAN/patchGAN.tex
\begin{figure*}[htbp]
	\includegraphics[width=\textwidth]{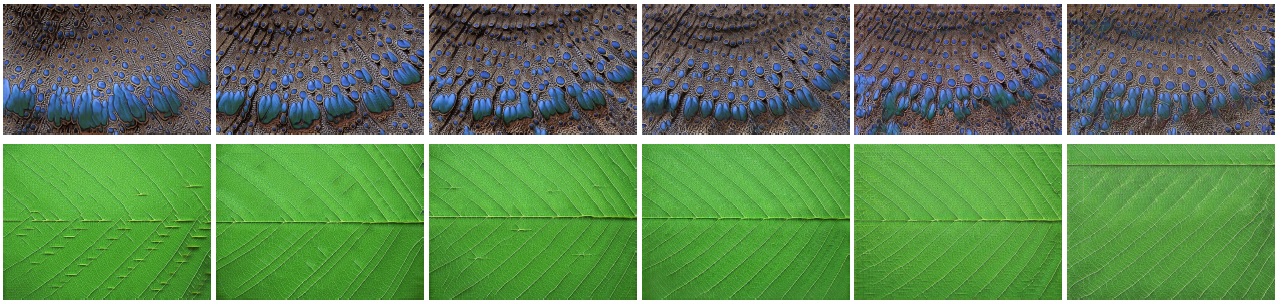}\\
	\vspace{1mm}
	\rotatebox{90}{\parbox[b]{1mm}{\vphantom{g}}}~
	\parbox{0.16\textwidth}{\centering 16$\times$16}~%
	\parbox{0.16\textwidth}{\centering 34$\times$34}~%
	\parbox{0.16\textwidth}{\centering 70$\times$70}~%
	\parbox{0.16\textwidth}{\centering 142$\times$142}~%
	\parbox{0.16\textwidth}{\centering 286$\times$286}~%
	\parbox{0.16\textwidth}{\centering 574$\times$574}%
	\caption{Comparison of PatchGANs with different receptive fields, ranging from 16$\times$16 to 574$\times$574, corresponding to 3 to 8 convolutional layers in the discriminator. Adding layers increases the receptive field (i.e., the patch size) of PatchGAN, which makes it possible for the discriminator to examine larger structures. However, as may be seen above, very large patch sizes seem to cause the discriminator to pay less attention to local details. We use a patch size of 142$\times$142 in our results.}
	\label{fig:patchGAN}
\end{figure*}

%% file: figures/stress1_crop/crop_expansion.tex
\begin{figure*}[htbp]
	\includegraphics[width=\textwidth]{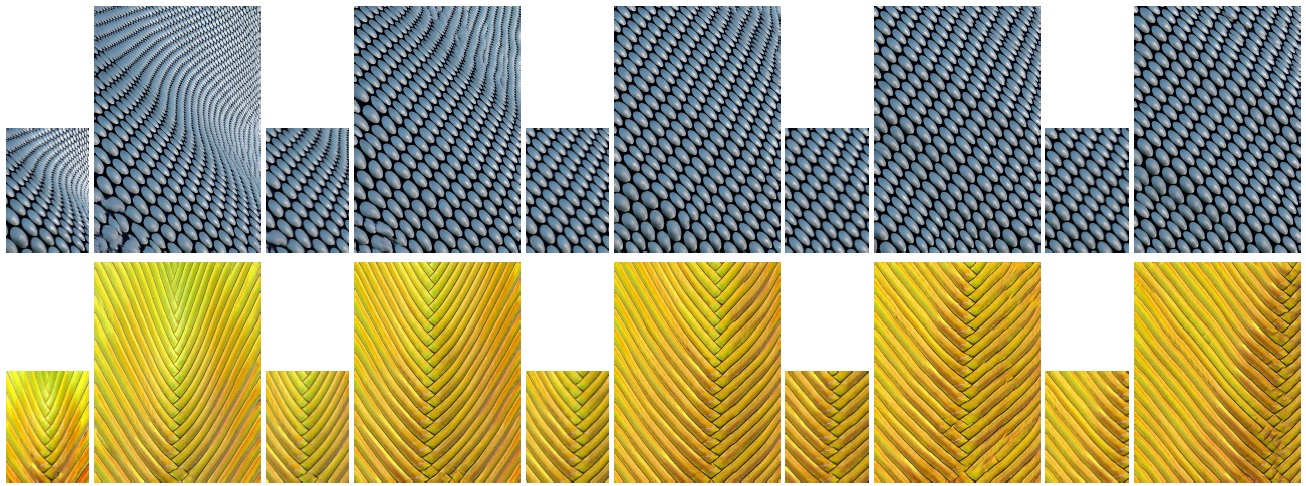}
	\caption{Stress test \#1. Given a source texture (leftmost column), we double its size using our method. Then we randomly crop a region of the same size as the source texture from the expansion result, and expand it again without any further training. The above crop-expansion cycle is repeated 4 times. We can see that the final result (rightmost column) is still very sharp and natural looking, attesting to the stability of our method.}
	\label{fig:stress1_crop}
\end{figure*}

%% file: figures/stress2_extreme/extreme.tex
\begin{figure*}[htbp]
	\includegraphics[width=0.95\textwidth]{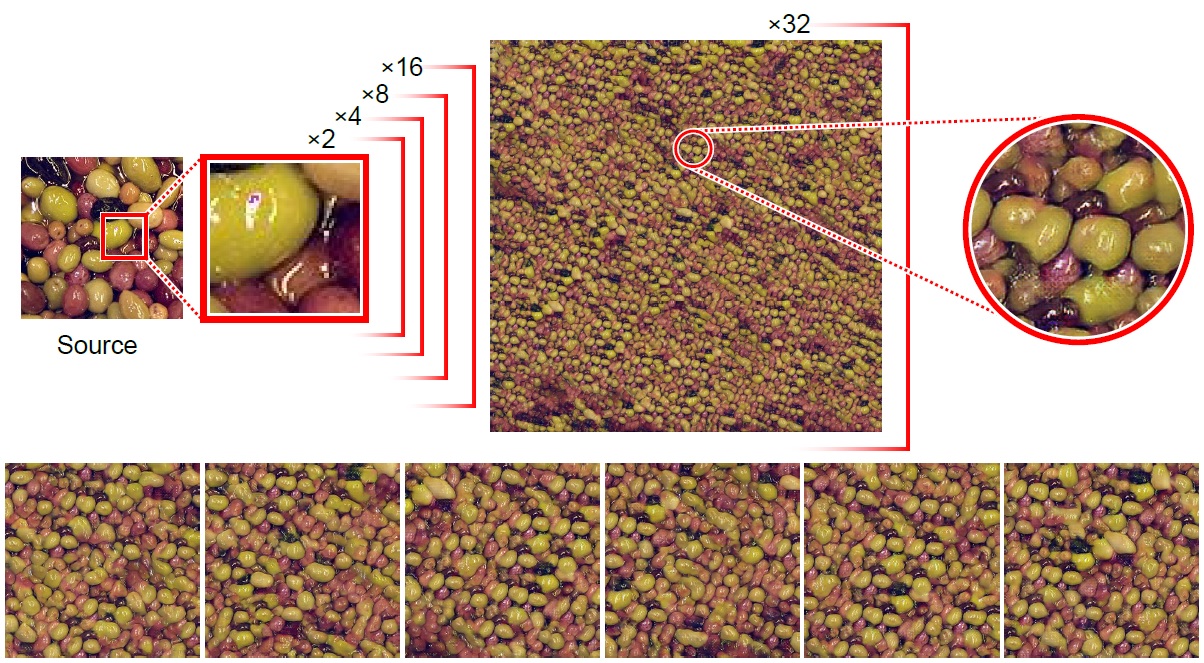}
	\caption{Extreme expansion. Having trained a generator on the source exemplar (left), we feed it with a small cropped texture block (64$\times$64 pixels), and feed the expanded result back into the generator. Five such cycles produce a 2048$\times$2048 result. Six different crops from this result are shown in the bottom row.}
	\label{fig:stress2_extreme}
\end{figure*}

%% file: summary.tex
\section{Summary}
\label{sec:summary}

We have presented an example-based texture synthesis method capable of expanding an exemplar texture, while faithfully preserving the global structures therein.
This is achieved by training a generative adversarial network, whose generator learns how to expand small subwindows of the exemplar to the larger texture windows containing them.
A variety of results demonstrate that, through such adversarial training, the generator is able to faithfully reproduce local patterns, as well as their global arrangements.
Although a dedicated generator must be trained for each exemplar, once it is trained, synthesis is extremely fast, requiring only a single feed-forward pass through the generator network.
The trained model is stable enough for repeated application, enabling generating diverse results of different sizes.

Training time is a limitation of our approach, although it is faster than previous GAN-based synthesis approaches. It would be useful to find a reliable stopping criterion for the training: at the moment, we train our models for 100,000 iterations, although in many cases the results no longer improve after 36,000 iterations or so.

\begin{figure}[htbp]
	\includegraphics[width=\columnwidth]{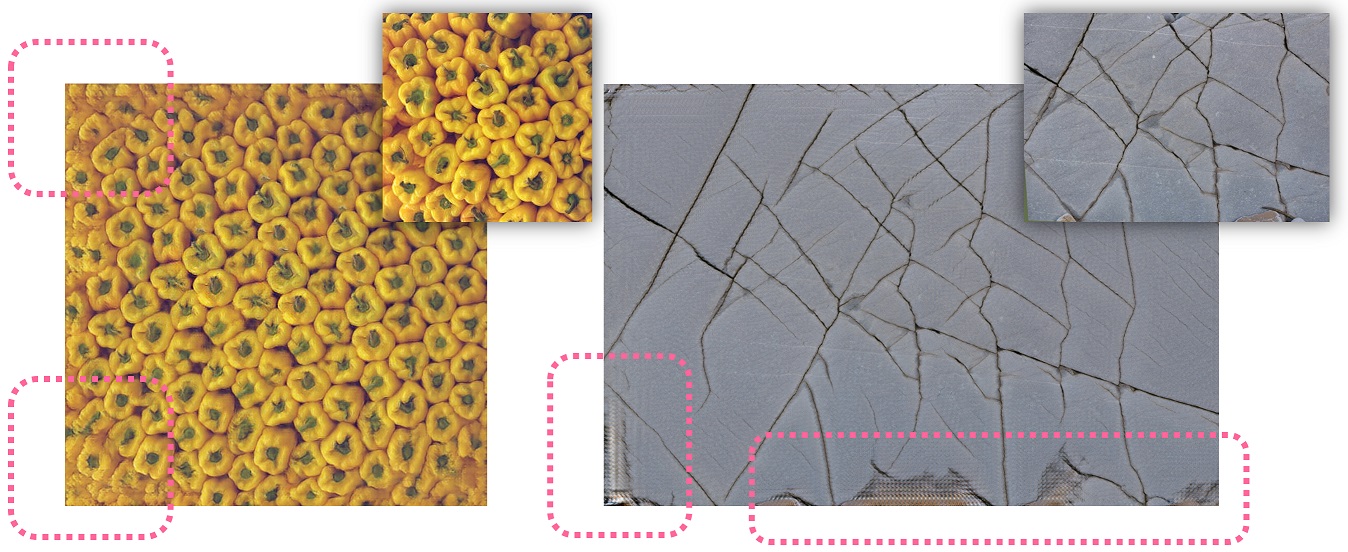}
	\caption{\yang{Artifacts in the border and corner regions.} }
	\label{fig:corner_issue}
\end{figure}

In terms of result quality, artifacts tend to emerge in the vicinity of borders and corners, as may be seen in \yang{Figure~\ref{fig:corner_issue}}. This may be attributed to fewer training examples in these areas, and possibly also related to the padding performed by the convolution layers.

\begin{figure}[htbp]
	\includegraphics[width=0.92\columnwidth]{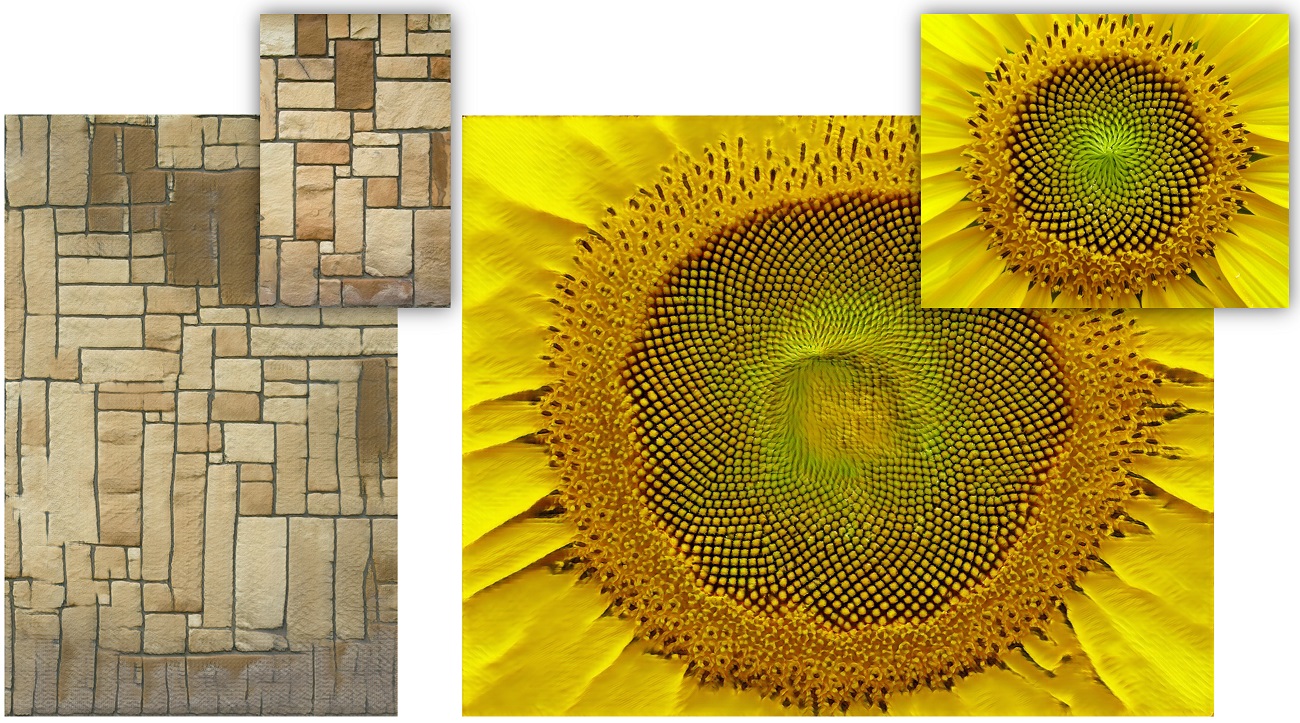}
	\caption{Failure cases of our method. For the stone tiles texture, our method failed to learn its large scale structure (left). While for the sunflower, our method failed to reproduce the singularity at the center (right).} 
	\label{fig:limitaions}
\end{figure}

Figure~\ref{fig:limitaions} shows two failure cases of our method. These failures may still be attributed to limited training examples. For example, for the stone tiles texture, all the tiles are quite large and distinct. So is the singularity at the center of the sunflower texture.
\new{In general, if the generator has not seen enough examples of a particular large scale structure or pattern during training, it cannot be expected to correctly reproduce and/or extend such structures during test time. The network does not learn some kind of a high-level representation of the texture; it only learns how to extend commonly occurring patterns.}
In the future, we would like to address this issue. It might be facilitated by training on multiple textures of the same class. With richer data we may possibly train a more powerful model for generalized texture synthesis tasks.




%% file: texexp.bbl

\begin{thebibliography}{31}


\ifx \showCODEN    \undefined \def \showCODEN     #1{\unskip}     \fi
\ifx \showDOI      \undefined \def \showDOI       #1{#1}\fi
\ifx \showISBNx    \undefined \def \showISBNx     #1{\unskip}     \fi
\ifx \showISBNxiii \undefined \def \showISBNxiii  #1{\unskip}     \fi
\ifx \showISSN     \undefined \def \showISSN      #1{\unskip}     \fi
\ifx \showLCCN     \undefined \def \showLCCN      #1{\unskip}     \fi
\ifx \shownote     \undefined \def \shownote      #1{#1}          \fi
\ifx \showarticletitle \undefined \def \showarticletitle #1{#1}   \fi
\ifx \showURL      \undefined \def \showURL       {\relax}        \fi
\providecommand\bibfield[2]{#2}
\providecommand\bibinfo[2]{#2}
\providecommand\natexlab[1]{#1}
\providecommand\showeprint[2][]{arXiv:#2}

\bibitem[\protect\citeauthoryear{Bergmann, Jetchev, and Vollgraf}{Bergmann
  et~al\mbox{.}}{2017}]%
        {Bergmann2017}
\bibfield{author}{\bibinfo{person}{Urs Bergmann}, \bibinfo{person}{Nikolay
  Jetchev}, {and} \bibinfo{person}{Roland Vollgraf}.}
  \bibinfo{year}{2017}\natexlab{}.
\newblock \showarticletitle{Learning Texture Manifolds with the Periodic
  Spatial {GAN}}.
\newblock \bibinfo{journal}{\emph{CoRR}}  \bibinfo{volume}{abs/1705.06566}
  (\bibinfo{year}{2017}).
\newblock


\bibitem[\protect\citeauthoryear{Darabi, Shechtman, Barnes, Goldman, and
  Sen}{Darabi et~al\mbox{.}}{2012}]%
        {Darabi12}
\bibfield{author}{\bibinfo{person}{Soheil Darabi}, \bibinfo{person}{Eli
  Shechtman}, \bibinfo{person}{Connelly Barnes}, \bibinfo{person}{Dan~B
  Goldman}, {and} \bibinfo{person}{Pradeep Sen}.}
  \bibinfo{year}{2012}\natexlab{}.
\newblock \showarticletitle{Image {M}elding: Combining Inconsistent Images
  using Patch-based Synthesis}.
\newblock \bibinfo{journal}{\emph{ACM Trans.~Graph. (Proc.~SIGGRAPH 2012)}}
  \bibinfo{volume}{31}, \bibinfo{number}{4} (\bibinfo{year}{2012}).
\newblock


\bibitem[\protect\citeauthoryear{Efros and Freeman}{Efros and Freeman}{2001}]%
        {Efros:2001}
\bibfield{author}{\bibinfo{person}{Alexei~A. Efros} {and}
  \bibinfo{person}{William~T. Freeman}.} \bibinfo{year}{2001}\natexlab{}.
\newblock \showarticletitle{Image quilting for texture synthesis and transfer}.
  In \bibinfo{booktitle}{\emph{Proc.~SIGGRAPH 2001}}.
  \bibinfo{pages}{341--346}.
\newblock


\bibitem[\protect\citeauthoryear{Efros and Leung}{Efros and Leung}{1999}]%
        {Efros99}
\bibfield{author}{\bibinfo{person}{Alexei~A. Efros} {and}
  \bibinfo{person}{Thomas~K. Leung}.} \bibinfo{year}{1999}\natexlab{}.
\newblock \showarticletitle{Texture Synthesis by Non-Parametric Sampling}.
\newblock \bibinfo{journal}{\emph{Proc.~ICCV '99}}  \bibinfo{volume}{2}
  (\bibinfo{year}{1999}), \bibinfo{pages}{1033--1038}.
\newblock


\bibitem[\protect\citeauthoryear{Gatys, Ecker, and Bethge}{Gatys
  et~al\mbox{.}}{2015a}]%
        {Gatys-texture-2015}
\bibfield{author}{\bibinfo{person}{L.A. Gatys}, \bibinfo{person}{A.S. Ecker},
  {and} \bibinfo{person}{M. Bethge}.} \bibinfo{year}{2015}\natexlab{a}.
\newblock \showarticletitle{Texture Synthesis Using Convolutional Neural
  Networks}.
\newblock \bibinfo{journal}{\emph{Advances in Neural Information Processing
  Systems}}  \bibinfo{volume}{28} (\bibinfo{date}{May} \bibinfo{year}{2015}),
  \bibinfo{pages}{262--270}.
\newblock
\urldef\tempurl%
\url{http://arxiv.org/abs/1505.07376}
\showURL{%
\tempurl}


\bibitem[\protect\citeauthoryear{Gatys, Ecker, and Bethge}{Gatys
  et~al\mbox{.}}{2015b}]%
        {Gatys2015c}
\bibfield{author}{\bibinfo{person}{Leon~A. Gatys},
  \bibinfo{person}{Alexander~S. Ecker}, {and} \bibinfo{person}{Matthias
  Bethge}.} \bibinfo{year}{2015}\natexlab{b}.
\newblock \showarticletitle{A Neural Algorithm of Artistic Style}.
\newblock \bibinfo{journal}{\emph{CoRR}}  \bibinfo{volume}{abs/1508.06576}
  (\bibinfo{date}{Aug} \bibinfo{year}{2015}).
\newblock
\urldef\tempurl%
\url{http://arxiv.org/abs/1508.06576}
\showURL{%
\tempurl}


\bibitem[\protect\citeauthoryear{Goodfellow, Pouget-Abadie, Mirza, Xu,
  Warde-Farley, Ozair, Courville, and Bengio}{Goodfellow et~al\mbox{.}}{2014}]%
        {Goodfellow2014}
\bibfield{author}{\bibinfo{person}{Ian~J. Goodfellow}, \bibinfo{person}{Jean
  Pouget-Abadie}, \bibinfo{person}{Mehdi Mirza}, \bibinfo{person}{Bing Xu},
  \bibinfo{person}{David Warde-Farley}, \bibinfo{person}{Sherjil Ozair},
  \bibinfo{person}{Aaron Courville}, {and} \bibinfo{person}{Yoshua Bengio}.}
  \bibinfo{year}{2014}\natexlab{}.
\newblock \showarticletitle{Generative {Adversarial} {Networks}}.
\newblock \bibinfo{journal}{\emph{CoRR}}  \bibinfo{volume}{abs/1406.2661}
  (\bibinfo{date}{June} \bibinfo{year}{2014}).
\newblock
\urldef\tempurl%
\url{https://arxiv.org/abs/1406.2661}
\showURL{%
\tempurl}


\bibitem[\protect\citeauthoryear{He, Zhang, Ren, and Sun}{He
  et~al\mbox{.}}{2016}]%
        {He2016}
\bibfield{author}{\bibinfo{person}{Kaiming He}, \bibinfo{person}{Xiangyu
  Zhang}, \bibinfo{person}{Shaoqing Ren}, {and} \bibinfo{person}{Jian Sun}.}
  \bibinfo{year}{2016}\natexlab{}.
\newblock \showarticletitle{Deep Residual Learning for Image Recognition}. In
  \bibinfo{booktitle}{\emph{Proc.~CVPR 2016}}.
\newblock


\bibitem[\protect\citeauthoryear{Hertzmann, Jacobs, Oliver, Curless, and
  Salesin}{Hertzmann et~al\mbox{.}}{2001}]%
        {Hertzmann01}
\bibfield{author}{\bibinfo{person}{Aaron Hertzmann},
  \bibinfo{person}{Charles~E. Jacobs}, \bibinfo{person}{Nuria Oliver},
  \bibinfo{person}{Brian Curless}, {and} \bibinfo{person}{David~H. Salesin}.}
  \bibinfo{year}{2001}\natexlab{}.
\newblock \showarticletitle{Image Analogies}.
\newblock \bibinfo{journal}{\emph{Proc.~SIGGRAPH 2001}} (\bibinfo{date}{August}
  \bibinfo{year}{2001}), \bibinfo{pages}{327--340}.
\newblock


\bibitem[\protect\citeauthoryear{Isola, Zhu, Zhou, and Efros}{Isola
  et~al\mbox{.}}{2016}]%
        {pix2pix2016-arxiv}
\bibfield{author}{\bibinfo{person}{Phillip Isola}, \bibinfo{person}{Jun{-}Yan
  Zhu}, \bibinfo{person}{Tinghui Zhou}, {and} \bibinfo{person}{Alexei~A.
  Efros}.} \bibinfo{year}{2016}\natexlab{}.
\newblock \showarticletitle{Image-to-Image Translation with Conditional
  Adversarial Networks}.
\newblock \bibinfo{journal}{\emph{CoRR}}  \bibinfo{volume}{abs/1611.07004}
  (\bibinfo{year}{2016}).
\newblock
\showeprint[arxiv]{1611.07004}
\urldef\tempurl%
\url{http://arxiv.org/abs/1611.07004}
\showURL{%
\tempurl}


\bibitem[\protect\citeauthoryear{Isola, Zhu, Zhou, and Efros}{Isola
  et~al\mbox{.}}{2017}]%
        {pix2pix2017-cvpr}
\bibfield{author}{\bibinfo{person}{Phillip Isola}, \bibinfo{person}{Jun-Yan
  Zhu}, \bibinfo{person}{Tinghui Zhou}, {and} \bibinfo{person}{Alexei~A
  Efros}.} \bibinfo{year}{2017}\natexlab{}.
\newblock \showarticletitle{Image-to-Image Translation with Conditional
  Adversarial Networks}. In \bibinfo{booktitle}{\emph{Proc.~CVPR 2017}}.
\newblock


\bibitem[\protect\citeauthoryear{Jetchev, Bergmann, and Vollgraf}{Jetchev
  et~al\mbox{.}}{2016}]%
        {Jetchev2016}
\bibfield{author}{\bibinfo{person}{Nikolay Jetchev}, \bibinfo{person}{Urs
  Bergmann}, {and} \bibinfo{person}{Roland Vollgraf}.}
  \bibinfo{year}{2016}\natexlab{}.
\newblock \showarticletitle{Texture Synthesis with Spatial Generative
  Adversarial Networks}.
\newblock \bibinfo{journal}{\emph{CoRR}}  \bibinfo{volume}{abs/1611.08207}
  (\bibinfo{year}{2016}).
\newblock


\bibitem[\protect\citeauthoryear{Johnson, Alahi, and Fei-Fei}{Johnson
  et~al\mbox{.}}{2016}]%
        {Johnson2016}
\bibfield{author}{\bibinfo{person}{Justin Johnson}, \bibinfo{person}{Alexandre
  Alahi}, {and} \bibinfo{person}{Li Fei-Fei}.} \bibinfo{year}{2016}\natexlab{}.
\newblock \showarticletitle{Perceptual Losses for Real-Time Style Transfer and
  Super-Resolution}. In \bibinfo{booktitle}{\emph{Proc.~ECCV 2016}},
  \bibfield{editor}{\bibinfo{person}{Bastian Leibe}, \bibinfo{person}{Jiri
  Matas}, \bibinfo{person}{Nicu Sebe}, {and} \bibinfo{person}{Max Welling}}
  (Eds.), Vol.~\bibinfo{volume}{Part II}. \bibinfo{pages}{694--711}.
\newblock


\bibitem[\protect\citeauthoryear{Kaspar, Neubert, Lischinski, Pauly, and
  Kopf}{Kaspar et~al\mbox{.}}{2015}]%
        {Kaspar:2015}
\bibfield{author}{\bibinfo{person}{A. Kaspar}, \bibinfo{person}{B. Neubert},
  \bibinfo{person}{D. Lischinski}, \bibinfo{person}{M. Pauly}, {and}
  \bibinfo{person}{J. Kopf}.} \bibinfo{year}{2015}\natexlab{}.
\newblock \showarticletitle{Self tuning texture optimization}.
\newblock \bibinfo{journal}{\emph{Computer Graphics Forum}}
  \bibinfo{volume}{34}, \bibinfo{number}{2} (\bibinfo{date}{May}
  \bibinfo{year}{2015}).
\newblock


\bibitem[\protect\citeauthoryear{Kingma and Ba}{Kingma and Ba}{2014}]%
        {Kingma2014-Adam}
\bibfield{author}{\bibinfo{person}{Diederik~P. Kingma} {and}
  \bibinfo{person}{Jimmy Ba}.} \bibinfo{year}{2014}\natexlab{}.
\newblock \showarticletitle{Adam: {A} Method for Stochastic Optimization}.
\newblock \bibinfo{journal}{\emph{CoRR}}  \bibinfo{volume}{abs/1412.6980}
  (\bibinfo{year}{2014}).
\newblock
\showeprint[arxiv]{1412.6980}
\urldef\tempurl%
\url{http://arxiv.org/abs/1412.6980}
\showURL{%
\tempurl}


\bibitem[\protect\citeauthoryear{Kwatra, Essa, Bobick, and Kwatra}{Kwatra
  et~al\mbox{.}}{2005}]%
        {Kwatra2005}
\bibfield{author}{\bibinfo{person}{Vivek Kwatra}, \bibinfo{person}{Irfan Essa},
  \bibinfo{person}{Aaron Bobick}, {and} \bibinfo{person}{Nipun Kwatra}.}
  \bibinfo{year}{2005}\natexlab{}.
\newblock \showarticletitle{Texture optimization for example-based synthesis}.
\newblock \bibinfo{journal}{\emph{ACM Trans.~Graph.}} \bibinfo{volume}{24},
  \bibinfo{number}{3 (Proc.~SIGGRAPH 2005)} (\bibinfo{year}{2005}),
  \bibinfo{pages}{795--802}.
\newblock


\bibitem[\protect\citeauthoryear{Kwatra, Sch\"{o}dl, Essa, Turk, and
  Bobick}{Kwatra et~al\mbox{.}}{2003}]%
        {Kwatra2003}
\bibfield{author}{\bibinfo{person}{Vivek Kwatra}, \bibinfo{person}{Arno
  Sch\"{o}dl}, \bibinfo{person}{Irfan Essa}, \bibinfo{person}{Greg Turk}, {and}
  \bibinfo{person}{Aaron Bobick}.} \bibinfo{year}{2003}\natexlab{}.
\newblock \showarticletitle{Graphcut textures: image and video synthesis using
  graph cuts}.
\newblock \bibinfo{journal}{\emph{ACM Trans.~Graph.}} \bibinfo{volume}{22},
  \bibinfo{number}{3 (Proc. SIGGRAPH 2003)} (\bibinfo{year}{2003}),
  \bibinfo{pages}{277--286}.
\newblock


\bibitem[\protect\citeauthoryear{Ledig, Theis, Huszar, Caballero, Aitken,
  Tejani, Totz, Wang, and Shi}{Ledig et~al\mbox{.}}{2016}]%
        {Ledig2016}
\bibfield{author}{\bibinfo{person}{Christian Ledig}, \bibinfo{person}{Lucas
  Theis}, \bibinfo{person}{Ferenc Huszar}, \bibinfo{person}{Jose Caballero},
  \bibinfo{person}{Andrew~P. Aitken}, \bibinfo{person}{Alykhan Tejani},
  \bibinfo{person}{Johannes Totz}, \bibinfo{person}{Zehan Wang}, {and}
  \bibinfo{person}{Wenzhe Shi}.} \bibinfo{year}{2016}\natexlab{}.
\newblock \showarticletitle{Photo-Realistic Single Image Super-Resolution Using
  a Generative Adversarial Network}.
\newblock \bibinfo{journal}{\emph{CoRR}}  \bibinfo{volume}{abs/1609.04802}
  (\bibinfo{year}{2016}).
\newblock
\showeprint[arxiv]{1609.04802}
\urldef\tempurl%
\url{http://arxiv.org/abs/1609.04802}
\showURL{%
\tempurl}


\bibitem[\protect\citeauthoryear{Lefebvre and Hoppe}{Lefebvre and
  Hoppe}{2006}]%
        {Lefebvre2006}
\bibfield{author}{\bibinfo{person}{Sylvain Lefebvre} {and}
  \bibinfo{person}{Hugues Hoppe}.} \bibinfo{year}{2006}\natexlab{}.
\newblock \showarticletitle{Appearance-space texture synthesis}.
\newblock \bibinfo{journal}{\emph{ACM Transactions on Graphics}}
  \bibinfo{volume}{25}, \bibinfo{number}{3 (Proc. SIGGRAPH 2006)}
  (\bibinfo{year}{2006}), \bibinfo{pages}{541--548}.
\newblock


\bibitem[\protect\citeauthoryear{Li and Wand}{Li and Wand}{2016}]%
        {Li2016}
\bibfield{author}{\bibinfo{person}{Chuan Li} {and} \bibinfo{person}{Michael
  Wand}.} \bibinfo{year}{2016}\natexlab{}.
\newblock \showarticletitle{Precomputed Real-Time Texture Synthesis with
  Markovian Generative Adversarial Networks}. In
  \bibinfo{booktitle}{\emph{Proc.~ECCV 2016}},
  \bibfield{editor}{\bibinfo{person}{Bastian Leibe}, \bibinfo{person}{Jiri
  Matas}, \bibinfo{person}{Nicu Sebe}, {and} \bibinfo{person}{Max Welling}}
  (Eds.), Vol.~\bibinfo{volume}{Part III}. \bibinfo{pages}{702--716}.
\newblock


\bibitem[\protect\citeauthoryear{Liu, Lin, and Hays}{Liu et~al\mbox{.}}{2004}]%
        {Liu04}
\bibfield{author}{\bibinfo{person}{Yanxi Liu}, \bibinfo{person}{Web-Chieh Lin},
  {and} \bibinfo{person}{James~H. Hays}.} \bibinfo{year}{2004}\natexlab{}.
\newblock \showarticletitle{Near-Regular Texture Analysis and Manipulation}.
\newblock \bibinfo{journal}{\emph{ACM Transactions on Graphics (Proceedings of
  SIGGRAPH 2004)}} \bibinfo{volume}{23}, \bibinfo{number}{3}
  (\bibinfo{year}{2004}).
\newblock


\bibitem[\protect\citeauthoryear{Rosenberger, Cohen-Or, and
  Lischinski}{Rosenberger et~al\mbox{.}}{2009}]%
        {Rosenberger:2009}
\bibfield{author}{\bibinfo{person}{Amir Rosenberger}, \bibinfo{person}{Daniel
  Cohen-Or}, {and} \bibinfo{person}{Dani Lischinski}.}
  \bibinfo{year}{2009}\natexlab{}.
\newblock \showarticletitle{Layered Shape Synthesis: Automatic Generation of
  Control Maps for Non-Stationary Textures}.
\newblock \bibinfo{journal}{\emph{ACM Trans.~Graph}} \bibinfo{volume}{28},
  \bibinfo{number}{5} (\bibinfo{date}{Dec.} \bibinfo{year}{2009}),
  \bibinfo{pages}{107:1--9}.
\newblock


\bibitem[\protect\citeauthoryear{Sendik and Cohen-Or}{Sendik and
  Cohen-Or}{2017}]%
        {Sendik2017}
\bibfield{author}{\bibinfo{person}{Omry Sendik} {and} \bibinfo{person}{Daniel
  Cohen-Or}.} \bibinfo{year}{2017}\natexlab{}.
\newblock \showarticletitle{Deep Correlations for Texture Synthesis}.
\newblock \bibinfo{journal}{\emph{ACM Trans. Graph.}} \bibinfo{volume}{36},
  \bibinfo{number}{5}, Article \bibinfo{articleno}{161} (\bibinfo{date}{July}
  \bibinfo{year}{2017}), \bibinfo{numpages}{15}~pages.
\newblock
\showISSN{0730-0301}
\urldef\tempurl%
\url{https://doi.org/10.1145/3015461}
\showDOI{\tempurl}


\bibitem[\protect\citeauthoryear{Simonyan and Zisserman}{Simonyan and
  Zisserman}{2014}]%
        {Simonyan2014}
\bibfield{author}{\bibinfo{person}{Karen Simonyan} {and}
  \bibinfo{person}{Andrew Zisserman}.} \bibinfo{year}{2014}\natexlab{}.
\newblock \showarticletitle{Very Deep Convolutional Networks for Large-Scale
  Image Recognition}.
\newblock \bibinfo{journal}{\emph{CoRR}}  \bibinfo{volume}{abs/1409.1556}
  (\bibinfo{year}{2014}).
\newblock
\showeprint[arxiv]{1409.1556}
\urldef\tempurl%
\url{http://arxiv.org/abs/1409.1556}
\showURL{%
\tempurl}


\bibitem[\protect\citeauthoryear{Ulyanov, Lebedev, Vedaldi, and
  Lempitsky}{Ulyanov et~al\mbox{.}}{2016}]%
        {Ulyanov2016}
\bibfield{author}{\bibinfo{person}{Dmitry Ulyanov}, \bibinfo{person}{Vadim
  Lebedev}, \bibinfo{person}{Andrea Vedaldi}, {and} \bibinfo{person}{Victor~S.
  Lempitsky}.} \bibinfo{year}{2016}\natexlab{}.
\newblock \showarticletitle{Texture Networks: Feed-forward Synthesis of
  Textures and Stylized Images}.
\newblock \bibinfo{journal}{\emph{CoRR}}  \bibinfo{volume}{abs/1603.03417}
  (\bibinfo{year}{2016}).
\newblock
\showeprint[arxiv]{1603.03417}
\urldef\tempurl%
\url{http://arxiv.org/abs/1603.03417}
\showURL{%
\tempurl}


\bibitem[\protect\citeauthoryear{Wei, Lefebvre, Kwatra, and Turk}{Wei
  et~al\mbox{.}}{2009}]%
        {Wei2009}
\bibfield{author}{\bibinfo{person}{Li-Yi Wei}, \bibinfo{person}{Sylvain
  Lefebvre}, \bibinfo{person}{Vivek Kwatra}, {and} \bibinfo{person}{Greg
  Turk}.} \bibinfo{year}{2009}\natexlab{}.
\newblock \showarticletitle{State of the Art in Example-based Texture
  Synthesis}. In \bibinfo{booktitle}{\emph{Eurographics 2009 State of The Art
  Reports}}. Eurographics.
\newblock


\bibitem[\protect\citeauthoryear{Wei and Levoy}{Wei and Levoy}{2000}]%
        {Wei2000}
\bibfield{author}{\bibinfo{person}{Li-Yi Wei} {and} \bibinfo{person}{Marc
  Levoy}.} \bibinfo{year}{2000}\natexlab{}.
\newblock \showarticletitle{Fast texture synthesis using tree-structured vector
  quantization}.
\newblock \bibinfo{journal}{\emph{Proc.~SIGGRAPH 2000}} (\bibinfo{year}{2000}),
  \bibinfo{pages}{479--488}.
\newblock


\bibitem[\protect\citeauthoryear{Wexler, Shechtman, and Irani}{Wexler
  et~al\mbox{.}}{2007}]%
        {Wexler07}
\bibfield{author}{\bibinfo{person}{Y. Wexler}, \bibinfo{person}{E. Shechtman},
  {and} \bibinfo{person}{M. Irani}.} \bibinfo{year}{2007}\natexlab{}.
\newblock \showarticletitle{Space-time completion of video}.
\newblock \bibinfo{journal}{\emph{Transactions on Pattern Analysis and Machine
  Intelligence (PAMI)}} \bibinfo{volume}{29}, \bibinfo{number}{3}
  (\bibinfo{year}{2007}), \bibinfo{pages}{463--476}.
\newblock


\bibitem[\protect\citeauthoryear{Zhang, Zhou, Velho, Guo, and Shum}{Zhang
  et~al\mbox{.}}{2003}]%
        {Zhang2003}
\bibfield{author}{\bibinfo{person}{Jingdan Zhang}, \bibinfo{person}{Kun Zhou},
  \bibinfo{person}{Luiz Velho}, \bibinfo{person}{Baining Guo}, {and}
  \bibinfo{person}{Heung-Yeung Shum}.} \bibinfo{year}{2003}\natexlab{}.
\newblock \showarticletitle{Synthesis of progressively-variant textures on
  arbitrary surfaces}.
\newblock \bibinfo{journal}{\emph{ACM Trans.~Graph.}} \bibinfo{volume}{22},
  \bibinfo{number}{3 (Proc.~SIGGRAPH 2003)} (\bibinfo{year}{2003}),
  \bibinfo{pages}{295--302}.
\newblock
\showISBNx{1-58113-709-5}


\bibitem[\protect\citeauthoryear{Zhou, Shi, Lischinski, Gong, Kopf, and
  Huang}{Zhou et~al\mbox{.}}{2017}]%
        {Zhou2017}
\bibfield{author}{\bibinfo{person}{Yang Zhou}, \bibinfo{person}{Huajie Shi},
  \bibinfo{person}{Dani Lischinski}, \bibinfo{person}{Minglun Gong},
  \bibinfo{person}{Johannes Kopf}, {and} \bibinfo{person}{Hui Huang}.}
  \bibinfo{year}{2017}\natexlab{}.
\newblock \showarticletitle{Analysis and Controlled Synthesis of Inhomogeneous
  Textures}.
\newblock \bibinfo{journal}{\emph{Computer Graphics Forum (Proc.~Eurographics
  2017)}} \bibinfo{volume}{36}, \bibinfo{number}{2} (\bibinfo{year}{2017}).
\newblock


\bibitem[\protect\citeauthoryear{Zhu, Park, Isola, and Efros}{Zhu
  et~al\mbox{.}}{2017}]%
        {CycleGAN2017}
\bibfield{author}{\bibinfo{person}{Jun{-}Yan Zhu}, \bibinfo{person}{Taesung
  Park}, \bibinfo{person}{Phillip Isola}, {and} \bibinfo{person}{Alexei~A.
  Efros}.} \bibinfo{year}{2017}\natexlab{}.
\newblock \showarticletitle{Unpaired Image-to-Image Translation using
  Cycle-Consistent Adversarial Networks}.
\newblock \bibinfo{journal}{\emph{CoRR}}  \bibinfo{volume}{abs/1703.10593}
  (\bibinfo{year}{2017}).
\newblock
\showeprint[arxiv]{1703.10593}
\urldef\tempurl%
\url{http://arxiv.org/abs/1703.10593}
\showURL{%
\tempurl}


\end{thebibliography}
